\documentclass[
superscriptaddress,
reprint,
showpacs,preprintnumbers,
amsmath,amssymb,
aps,
prl,
longbibliography
]{revtex4-2}
\usepackage{color}
\usepackage{graphicx}
\usepackage{bm}
\usepackage{mathrsfs}
\usepackage{dsfont}
\usepackage{ulem}
\usepackage[unicode=true, breaklinks=false, pdfborder={0 0 1}, backref=false, colorlinks=true, linkcolor=blue, citecolor=blue]{hyperref}
\usepackage{physics}

\begin{document}
\title{Room-Temperature Electrical Readout of Spin Defects in van der Waals Materials}

\author{Shihao Ru}
\affiliation{School of Electrical and Electronic Engineering, Nanyang Technological University, Singapore}
\affiliation{Division of Physics and Applied Physics, School of Physical and Mathematical Sciences, Nanyang Technological University, Singapore 637371, Singapore}
\affiliation{Centre for Quantum Technologies, Nanyang Technological University, Singapore 117543, Singapore}
\affiliation{National Centre for Advanced Integrated Photonics, Nanyang Technological University, Singapore 639798, Singapore}

\author{Liheng An}
\affiliation{Division of Physics and Applied Physics, School of Physical and Mathematical Sciences, Nanyang Technological University, Singapore 637371, Singapore}

\author{Haidong Liang}
\affiliation{Centre for Ion Beam Applications, Department of Physics, National University of Singapore, Singapore 117542, Singapore}

\author{Zhengzhi Jiang}
\affiliation{Department of Chemistry, National University of Singapore, Singapore 117543, Singapore}

\author{Zhiwei Li}
\affiliation{School of Electrical and Electronic Engineering, Nanyang Technological University, Singapore}

\author{Xiaodan Lyu}
\affiliation{Division of Physics and Applied Physics, School of Physical and Mathematical Sciences, Nanyang Technological University, Singapore 637371, Singapore}

\author{Feifei Zhou}
\affiliation{College of Metrology Measurement and Instrument, China Jiliang University, Hangzhou, 310018, China}

\author{Hongbing Cai}
\affiliation{Hefei National Laboratory for Physical Sciences at the Microscale, University of Science and Technology of China, Hefei Anhui 230026, China}

\author{Yuzhe Yang}
\affiliation{Division of Physics and Applied Physics, School of Physical and Mathematical Sciences, Nanyang Technological University, Singapore 637371, Singapore}

\author{Ruihua He}
\affiliation{Division of Physics and Applied Physics, School of Physical and Mathematical Sciences, Nanyang Technological University, Singapore 637371, Singapore}
\affiliation{School of Biological Sciences, Nanyang Technological University, Singapore 637551, Singapore}

\author{Robert Cernansky}
\affiliation{Institute for Quantum Optics, Ulm University, Albert-Einstein-Allee 11, 89081 Ulm, Germany}

\author{Edwin Hang Tong Teo}
\affiliation{School of Electrical and Electronic Engineering, Nanyang Technological University, Singapore}

\author{Manas Mukherjee}
\affiliation{Centre for Quantum Technologies, National University of Singapore, Singapore 117543, Singapore}

\author{Andrew A. Bettiol}
\affiliation{Centre for Ion Beam Applications, Department of Physics, National University of Singapore, Singapore 117542, Singapore}

\author{Jesus Z\'u\~niga-Perez}
\affiliation{Division of Physics and Applied Physics, School of Physical and Mathematical Sciences, Nanyang Technological University, Singapore 637371, Singapore}
\affiliation{Majulab, International Research Laboratory IRL 3654, CNRS, Université Côte d’Azur, Sorbonne Université, National University of Singapore, Nanyang Technological University, Singapore, Singapore}

\author{Fedor Jelezko}
\email[]{fedor.jelezko@umi-ulm.de}
\affiliation{Institute for Quantum Optics, Ulm University, Albert-Einstein-Allee 11, 89081 Ulm, Germany}

\author{Weibo Gao}
\email[]{wbgao@ntu.edu.sg}
\affiliation{School of Electrical and Electronic Engineering, Nanyang Technological University, Singapore}
\affiliation{Division of Physics and Applied Physics, School of Physical and Mathematical Sciences, Nanyang Technological University, Singapore 637371, Singapore}
\affiliation{Centre for Quantum Technologies, Nanyang Technological University, Singapore 117543, Singapore}
\affiliation{National Centre for Advanced Integrated Photonics, Nanyang Technological University, Singapore 639798, Singapore}

\date{\today}

\begin{abstract}
Negatively charged boron vacancy ($\mathrm{V_B^-}$) in hexagonal boron nitride (hBN) is the most extensively studied room-temperature quantum spin system in two-dimensional (2D) materials. Nevertheless, the current effective readout of $\mathrm{V_B^-}$ spin states is carried out by systematically optical methods. This limits their exploitation in compact and miniaturized quantum devices, which would otherwise hold substantial promise to address quantum sensing and quantum information tasks. In this study, we demonstrated a photoelectric spin readout technique for $\mathrm{V_B^-}$ spins in hBN. The observed photocurrent signals stem from the spin-dependent ionization dynamics of boron vacancies, mediated by spin-dependent non-radiative transitions to a metastable state. We further extend this electrical detection technique to enable the readout of dynamical decoupling sequences, including the Carr-Purcell-Meiboom-Gill (CPMG) protocols, and of nuclear spins via electron-nuclear double resonance. These results provide a pathway toward on-chip integration and real-field exploitation of quantum functionalities based on 2D material platforms.

\end{abstract}

\maketitle

The study of quantum technologies based on 2D materials is gaining increasing attention, particularly in the field of quantum sensing \cite{RN1,RN2,RN3}.
This interest is fueled by the exceptional physical and electronic properties of 2D materials, including the high carrier mobility of graphene \cite{RN4}, the direct bandgap nature of transition metal dichalcogenides, and the diverse quantum spin defects \cite{RN5,RN6,RN7,RN8} in hexagonal boron nitride (hBN). Notably, the negatively charged boron vacancy ($\mathrm{V_B^-}$) defects in hBN exhibit high responsiveness to external environmental factors like magnetic fields \cite{RN9, RN10}, strain or electric fields \cite{RN11}, and temperature \cite{RN12, RN13}, making them ideal candidates for high-sensitivity quantum sensors \cite{RN1, RN14}. First, the layered structure of hBN supports integration with nanophotonic devices \cite{RN15} and offers an ideal platform for quantum sensing at the atomic scale \cite{RN16}, which makes it well-suited for exploring spatially varying magnetic fluctuations \cite{RN17} or local spin dynamics in magnetic solid-state materials \cite{RN18, RN19}. Furthermore, these materials are highly compatible with existing silicon-based electronic technologies, enabling seamless integration and advancing the practical application of quantum technology. Finally, in contrast to traditional solid-state platforms, the atomic thickness and mechanical flexibility of hBN enable their integration into advanced probe architectures, such as atomic force microscopy tips or single-electron transistors \cite{RN16, RN20}, which are capable of high-resolution optical and electrical scanning. These properties position hBN as an ideal candidate for miniaturized sensing systems. 

\begin{figure*}[htpt]
  \centering
  \includegraphics[width=0.85\textwidth]{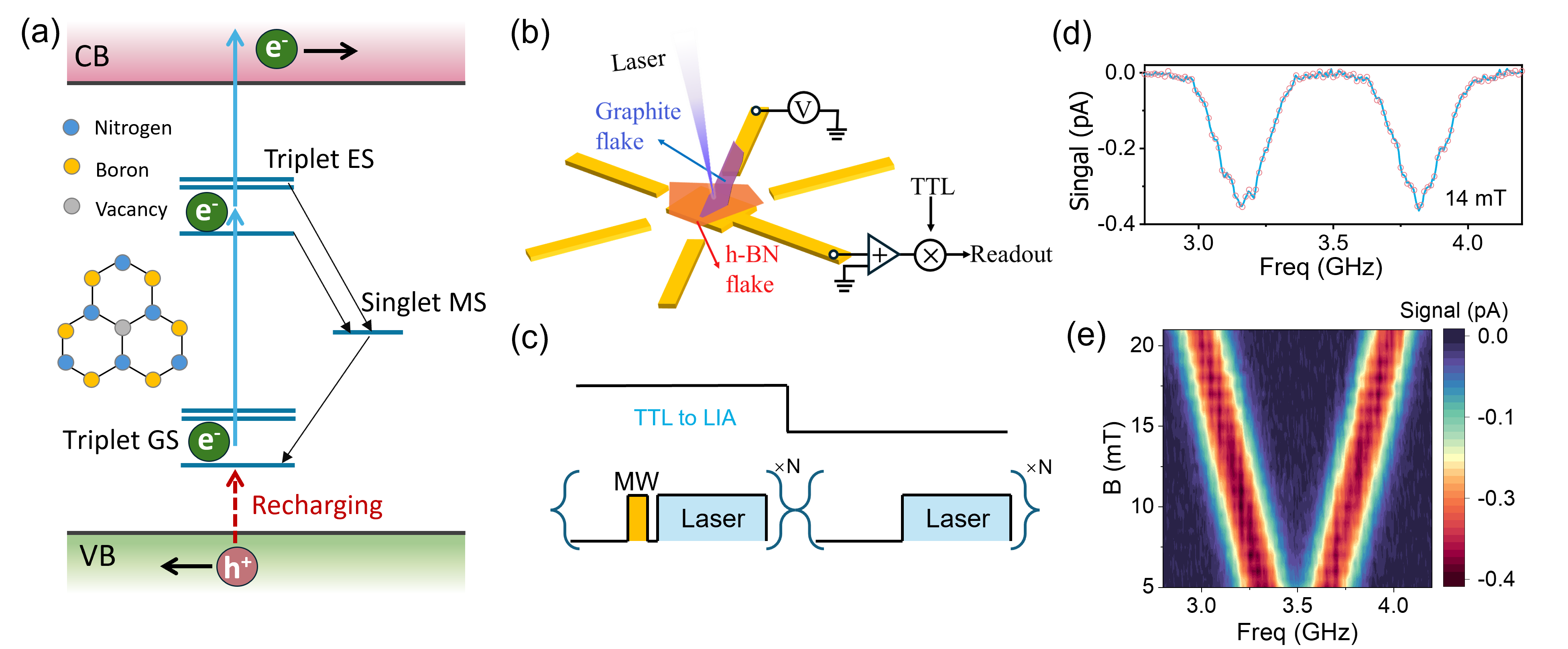}
  \caption{Schematic illustration and basic results of PDMR.
(a) Illustration of the charge dynamics: (1) Single-photon excitation promotes electrons to the excited state (ES), which can relax to the ground state (GS) or metastable state (MS) via spin-dependent intersystem crossing; (2) additional photon absorption elevates electrons to the conduction band; (3) charge repumping and carrier separation generate the photoelectric signal.
(b) Experimental setup for photoelectric readout.
(c) Timing sequence for pulsed PDMR. The signal demodulated using the TTL reference captures the microwave-induced signal.
(d) Representative PDMR spectrum at 14 mT.
(e) Magnetic field–dependent PDMR spectral map of the$m_s=\pm1$ states at $-2.3$ V bias and 2 mW laser power.}
\label{fig: figure1}
\end{figure*}

To fully leverage the above potentials, there is growing interest in developing compact magnetometers. 
The electron spin state of $\mathrm{V_B^-}$ centers is conventionally read out using optically detected magnetic resonance (ODMR), which relies on confocal microscopy to collect spin-dependent photoluminescence (PL). However, this method requires precise alignment of photon collection optics and bulky instrumentation, making it inherently incompatible with scalable integration into ultrathin 2D material-based electronic chips. To tackle this problem, a promising approach can be the photoelectric detection of magnetic resonance (PDMR), where a spin-dependent photoelectric current is generated under laser illumination. PDMR eliminates the need for photon collection optics by directly transducing spin states into measurable electrical signals. Previous studies have demonstrated electrical spin state readout in systems such as spin dopants in silicon \cite{RN21, RN22}, quantum dots \cite{RN23, RN24}, nitrogen-vacancy centers (both single defects \cite{RN25, RN26} and ensembles \cite{RN27, RN28}) in diamond, and silicon vacancies in 4H-SiC \cite{RN29} or 6H-SiC \cite{RN30}.

In this work, we demonstrate room-temperature electrical readout of spins of $\mathrm{V_B^-}$ in hexagonal boron nitride. We begin by detailing the implementation of PDMR on $\mathrm{V_B^-}$ centers in hBN. The device architecture incorporates a trilayer graphite top contact and gold bottom electrode, enabling both bias voltage application and photocurrent extraction. We achieve PDMR for $\mathrm{V_B^-}$ ensemble spins under various external magnetic fields. The observed photocurrent originates from two-photon ionization and spin dependent charge dynamics, which involves spin-dependent non-radiative intersystem crossing between the excited states and a metastable state. We also demonstrate coherent spin manipulation, including Rabi oscillations and dynamical decoupling protocols, as well as the electrical detection of coupled nuclear spins via electron-nuclear double resonance. 

\textit{Principle of photoelectric detection magnetic resonance.}
We begin by introducing the energy level structure of $\mathrm{V_B^-}$ defects. The $\mathrm{V_B^-}$, consisting of a missing boron atom lattice site in the AB-stacked hBN, offers stable deep-level energy states within a wide-bandgap host and exhibits spin-dependent intersystem crossing (ISC) \cite{RN7}. The $\mathrm{V_B^-}$ defect exhibits a spin-triplet ground state, a spin-triplet excited state, and a non-radiative singlet metastable state. At room temperature, the g-factors for the ground and excited states are approximately $2$, with zero-field splitting of $\approx3472$ MHz and $\approx2117$ MHz, respectively \cite{RN31, RN32, RN33}. The singlet metastable state provides nonradiative and spin-dependent ISC relaxation, allowing for optical spin state initialization and readout at room temperature. 

The concept of photoelectric detection is based on the intrinsic charge dynamics of the system, which proceed in three steps, as illustrated in Fig. \ref{fig: figure1}(a). First, the $\mathrm{V_B^-}$ defects can absorb a photon and transition from ground state to excited state. The excited states of $\mathrm{V_B^-}$ lay near the conduction band, enabling the electron to either decay radiatively or absorb a second photon and be promoted into the conduction band. It is important to note that while radiative decay rates from excited states to ground states are nearly equal and spin-independent, the nonradiative decay rates governed by the ISC through a metastable state are spin-dependent, forming the basis for spin signal readout under non-resonant excitation. Within one cycle, the negative charge state transitions to a neutral state when an excess electron is promoted to the conduction band through a second optical excitation. Finally, the system captures another electron, returning to its equilibrium charge state through a process known as charge repumping. This charge repumping process is hypothesized to be enhanced under laser pumping. 

In the PDMR experiment, the hBN flake with $\mathrm{V_B^-}$ defects was placed on a bottom electrode, and a trilayer graphite sheet was added on top as the bias contact. The bottom electrode was connected to a low-noise preamplifier and lock-in amplifier (LIA) for photoelectric signal detection.
A schematic of the device structure is illustrated in Fig. \ref{fig: figure1}(b). Figure \ref{fig: figure1}(c) shows a typical pulsed microwave (MW) timing sequence employed to identify the resonant MW frequency.
A 457 nm laser optically initializes the spin and charge states of $\mathrm{V_B^-}$ defects. Spin control is achieved via an MW antenna on a custom PCB, generating Rabi frequencies exceeding 40 MHz. The resulting photocurrent is converted to voltage by the preamplifier and demodulated by a LIA.
The TTL signal from a pulse generator served as the lock-in reference and captured the photocurrent difference with/without MW resonance.

Pulsed PDMR measurements followed the timing sequence in Fig. \ref{fig: figure1}(c), using a 200 ns MW pulse, a 2.1 $\mu$s laser pulse, and a 1.86 kHz TTL reference 1. To enhance the signal-to-noise ratio (SNR), the LIA used a 300 ms integration time, and the signals were averaged over a 1-second buffer.
The resulting pulsed PDMR spectrum, plotted as the photocurrent variation versus MW frequency under an out-of-plane magnetic field of 14 mT, is presented in Fig. \ref{fig: figure1}(d).
Two distinct peaks appear at 3.162 GHz and 3.817 GHz, corresponding to the $m_s=-1 (+1)\leftrightarrow m_s=0$ spin transitions of $\mathrm{V_B^-}$ defects in hBN, respectively.
To further investigate the magnetic field response, field–dependent measurements were performed by sweeping the field from 5 to 21 mT along the c-axis of $\mathrm{V_B^-}$ defects. Figure \ref{fig: figure1}(e) shows a broad range pulsed PDMR spectrum as a function of the magnetic field. The extracted ground-state g-factor is approximately 2, in agreement with values reported from optical readout methods \cite{RN31}.
\begin{figure}[htbp]
  \centering
  \includegraphics[width=0.5\textwidth]{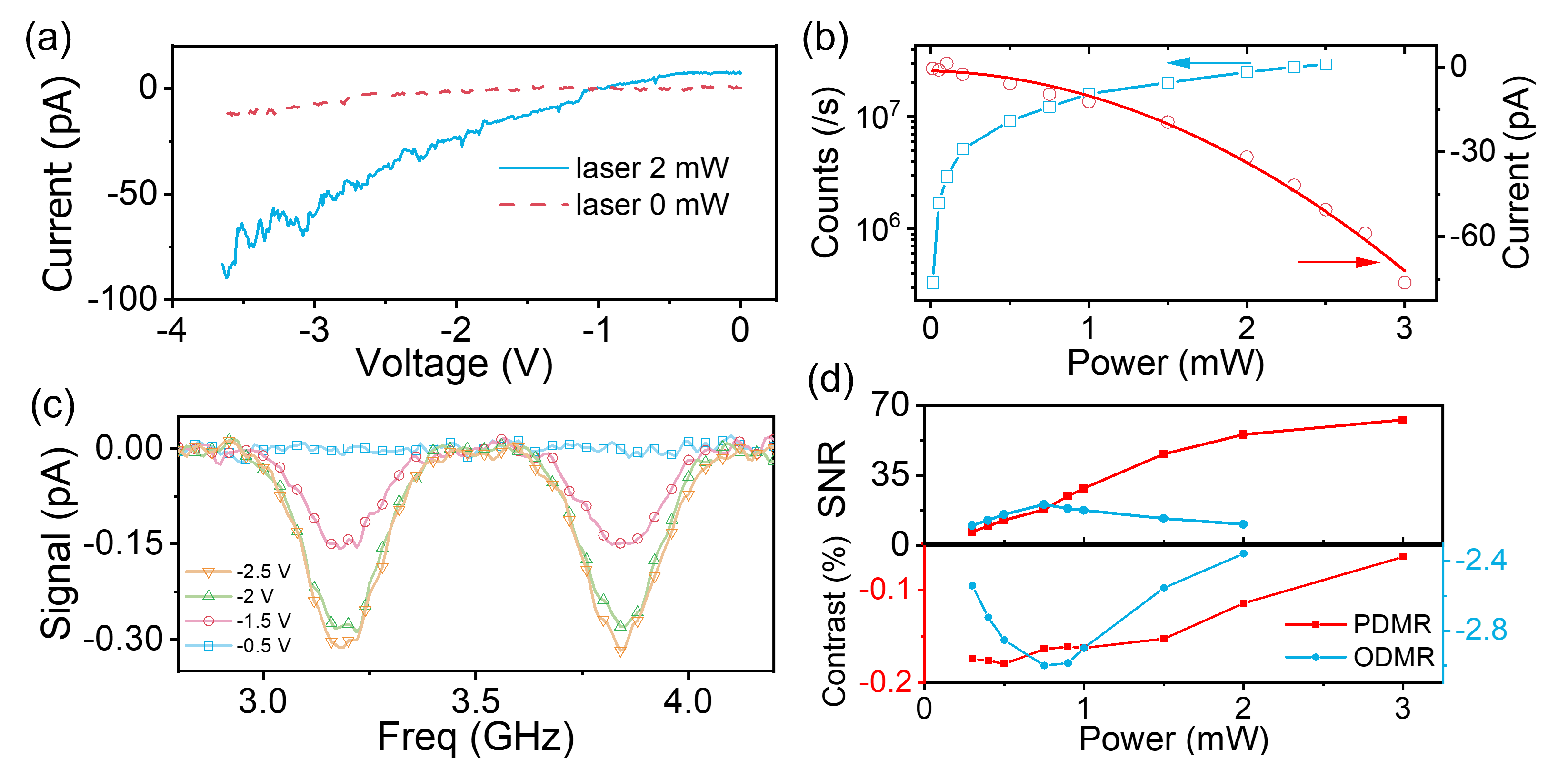}
  \caption{Dependence of PDMR signal and photocurrent on bias voltage and laser power.
(a) Photocurrent as a function of applied bias voltage under 2 mW laser excitation.
(b) Photoluminescence counts (blue squares) and photocurrent (red circles) versus laser power at a fixed bias of 2.3 V. The red curve is a quadratic fit to the photocurrent: $y=y_0+ax+bx^2$.
(c) PDMR spectra at 14 mT under varying bias voltages.
(d) PDMR contrast and SNR (red squares) and ODMR contrast and SNR (blue circles) as functions of laser power at 2.3 V bias.}
  \label{fig: figure2}
\end{figure}
Both demodulated outputs represent the magnitude signals from the LIA and are strictly positive. Since the $m_s=0$ state is optically bright, we define the PDMR signal to have negative signals.
Additionally, our setup allows for clearly resolving PDMR features within nuclear sublevels with a good SNR, Fig. \ref{fig: figure1}(d), via MW frequency scanning. This capability is essential for fast multi-frequency quantum control and enables applications such as electron-nuclear double resonance.

\textit{The effects of bias voltage and laser power.}
To examine the effects of laser power and bias voltage on the PDMR signal, Fig. \ref{fig: figure2} compares results under varying conditions. In Fig. \ref{fig: figure2}(a), the blue curve shows a photocurrent under 2 mW laser excitation, while the red dashed line indicates the leakage current without illumination, confirming laser-induced carrier generation. Fig. \ref{fig: figure2}(b) shows that both PL counts and photocurrent increase with laser power at a fixed 2.3 V bias. PL measurements (blue squares/curve) were limited to below 2.5 mW due to detector saturation, with no clear sign of PL saturation in that range.
Higher photon count rates allow the optical readout to approach the single-shot noise limit more rapidly. The measured photocurrent as a function of laser power (red circles) is well fitted by the function of $y=y_0+ax+bx^2$, with fitting parameters $a=-1.481$ $\mathrm{pA/mW}$ and $b=-7.363$ $\mathrm{pA/mW^2}$. The dominant quadratic term indicates a significant contribution from a two-photon absorption process, which is similar to previous reports on silicon vacancies \cite{RN29} in 4H-SiC and NV centers \cite{RN25, RN26, RN27, RN28} in diamond. Pulsed PDMR signals were measured at 14 mT under 2 mW laser excitation while varying the bias voltage, with results shown in Fig. \ref{fig: figure2}(c).
PDMR signal increases with bias voltage but saturates beyond $~\sim2$ V, showing minimal change at higher voltages. Figure \ref{fig: figure2}(d) compares pulsed ODMR and PDMR signals under varying laser powers ({Fig. S6 in \cite{Supp_info} for a magnified view}). As shown in the bottom panel, the PDMR contrast decreases at higher powers, whereas the ODMR first increases and then decreases as the power increases. The electrical SNR (top panel) continues to increase with laser power, eventually surpassing the optical SNR without signs of saturation. To balance performance and device stability, most measurements were performed at 2.3 V bias and 2 mW laser power \cite{Supp_info}.

\begin{figure}[htpt]
  \centering
  \includegraphics[width=0.5\textwidth]{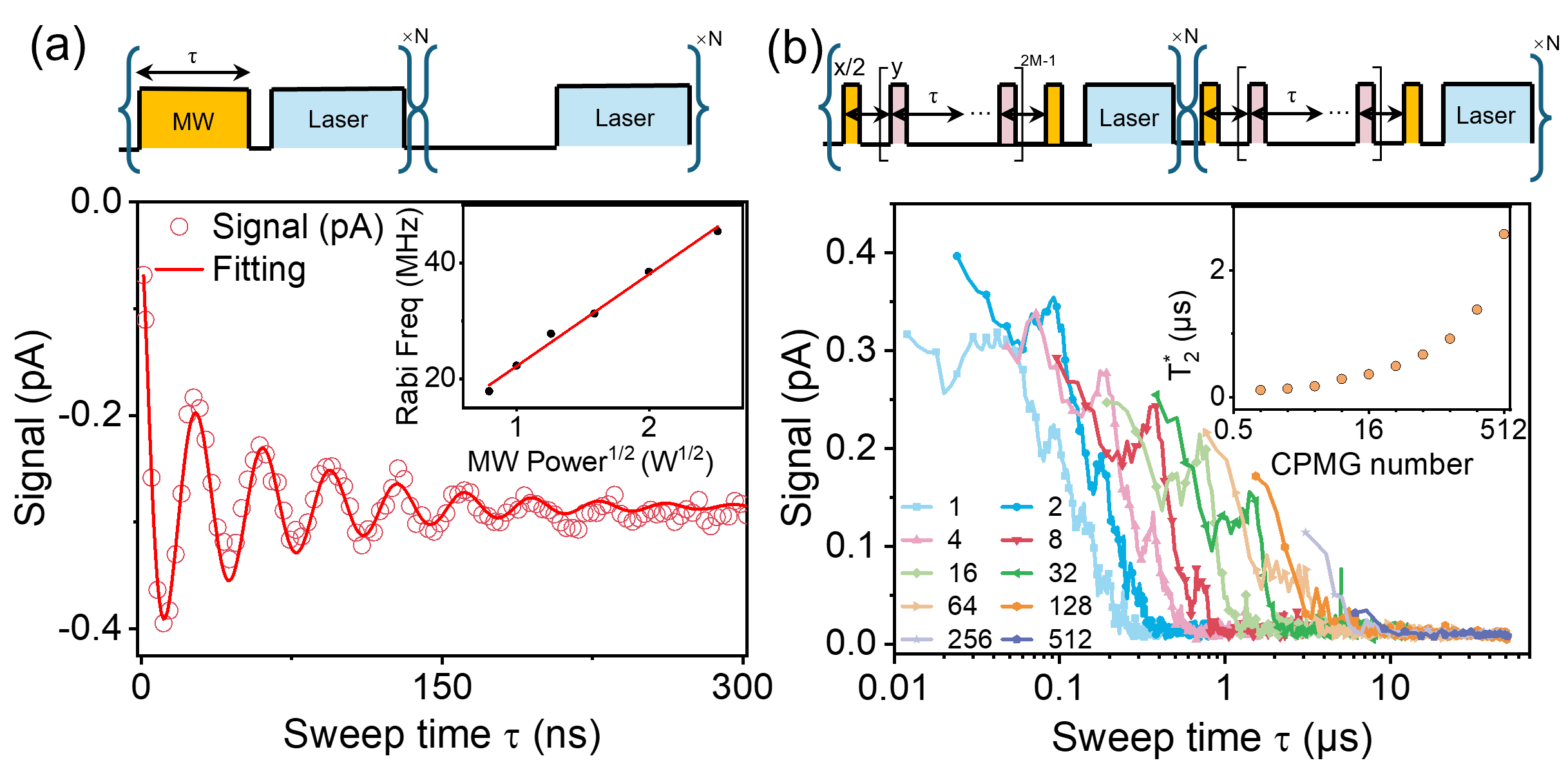}
  \caption{Coherent measurements of Rabi oscillations and dynamical decoupling of boron vacancies.
(a) The top panel shows the timing sequence for Rabi measurements. Red dots indicate photoelectric readout data, and the red line is the corresponding fit. The inset displays the Rabi frequency as a function of MW driving strength.
(b) The top panel shows the timing sequence for the CPMG measurement, with the first half labeled as the signal and the second as the reference. The bottom panel presents experimental results for CPMG sequences with 1 to 512 $\pi$ pulses. The inset shows the extracted coherence time versus the number of $\pi$ pulses on a log2 scale.}
  \label{fig: figure3}
\end{figure}
\textit{Electric readout of electron spin control and dynamics decoupling.}
The photoelectric method can be further developed to enable readout and demonstration of coherent electron spin control. To demonstrate its versatility, we present photoelectric readout for Rabi oscillations and dynamical decoupling sequences, including the CPMG sequences \cite{RN34, RN35}. These coherence measurements are fundamental for advanced quantum control protocols and provide valuable insight into nuclear spin interactions.

The timing sequence for Rabi measurements, shown in the top panel of Fig. \ref{fig: figure3}(a), closely follows the pulsed PDMR scheme depicted in Fig. \ref{fig: figure1}(d), with the MW frequency fixed at the central resonance of 3.165 GHz, corresponding to the $m_s=-1\leftrightarrow m_s=0$ transition at 14 mT. The MW pulse duration varies along the $x$-axis to observe Rabi oscillations, while the lock-in detection scheme remains identical to that used in the pulsed PDMR measurements (Fig. \ref{fig: figure1}).
The measured Rabi PDMR signal obtained at 14 mT and a MW drive of 38 dBm is shown as red dots in the bottom panel of Fig. \ref{fig: figure3}(a), with the corresponding fit represented by the red curve.
Rabi oscillations were modeled well as a sum of three exponentially damped sinusoid, $y(t)=y_0+\sum_{i=0}^2 A_i\exp(-t/\tau_i)\sin(2\pi(t-x_i)/\omega_i)$, with each term defined by amplitude $A_i$, phase offset $x_i$, oscillation period $\omega_i$ and decay constant $\tau_i$. The dominant component captures the MW Rabi rate (for $m_I=0$), while the other two provide smaller corrections associated with detuning ($m_I=\pm1$). Varying the drive power reveals a linear dependence of the Rabi frequency on the square root of the MW power [Fig.~\ref{fig: figure3}(a) inset].

Dynamics decoupling plays a crucial role in extending the coherence time of spin qubits, exploring nuclear spin interaction mechanisms, and enabling further precise nuclear spin manipulation. Utilizing the measured Rabi oscillation results, we set the Rabi frequency to 41.67 MHz, corresponding to a $\pi$-pulse duration of 12 ns, as the foundation for the subsequent CPMG measurements.
As shown in the upper panel of Fig. \ref{fig: figure3}(b), the CPMG timing sequences consist of two components: a signal sequence and a reference sequence. In the signal sequence, the spin state is first initiated to $m_s=0$ by optical pumping. A MW $\pi/2$-pulse along the $x$-axis then rotates the spin into a coherent superposition, i.e., the, the, the diagonal state of $m_s=0$ and $m_s=-1$.
This is followed by a series of MW $\pi$ pulses along the $y$-axis, and the sequence concludes with a final $\pi/2$ pulse along the $x$-axis. The reference sequence differs only in the final MW pulse, replacing the concluding $\pi/2$-pulse with a $-\pi/2$-pulse along the $x$-axis. It is noted that, in the CPMG measurements, the number of repetitions N was adjusted to keep the resulting TTL reference 1 input to the LIA within the range of 1.7–2 kHz, a range selected to optimize the SNR.
CPMG-PDMR were performed for sequences ranging from CPMG-1 to CPMG-512, with the results presented in the bottom panel of Fig. \ref{fig: figure3}(b).
Notably, several noticeable dips in the $\mathrm{V_B^-}$ fluorescence signal at specific evolution times are observed in the CPMG results. These features are attributed to coherent interactions between the electron spin and nearby polarized nitrogen nuclear spins, indicating a hyperfine coupling frequency of approximately 45.5 MHz. Such dips have traditionally been associated with electron-nuclear spin interactions and are typically revealed using extended dynamical decoupling sequences, which enhance electron-nuclear coupling by prolonging their interaction time. With the application of the CPMG-512 sequence, the spin coherence time can be extended from 0.13 to 2.57 $\mu$s, as shown in the inset of Fig. \ref{fig: figure3}(b).
In comparison, the XY-8 protocol (Fig. {S7} in \cite{Supp_info}) is less effective in preserving coherence time for $\mathrm{V_B^-}$ in hBN. This discrepancy may be due to spin-locking effects and strong coupling with surrounding nitrogen nuclear spins, whereas CPMG sequences are less susceptible to pulse errors, as previously reported \cite{RN36}.

\textit{Photoelectric readout of nuclear spins}
Finally, we demonstrate the electrical detection of nearby nitrogen nuclear spins associated with $\mathrm{V_B^-}$ defects in hBN using PDMR-based ENDOR measurements. In the experiment, a wire antenna was employed to deliver the radio frequency (RF) signals required for nuclear spin manipulation.
 \begin{figure}[htpt]
  \centering
  \includegraphics[width=0.4\textwidth]{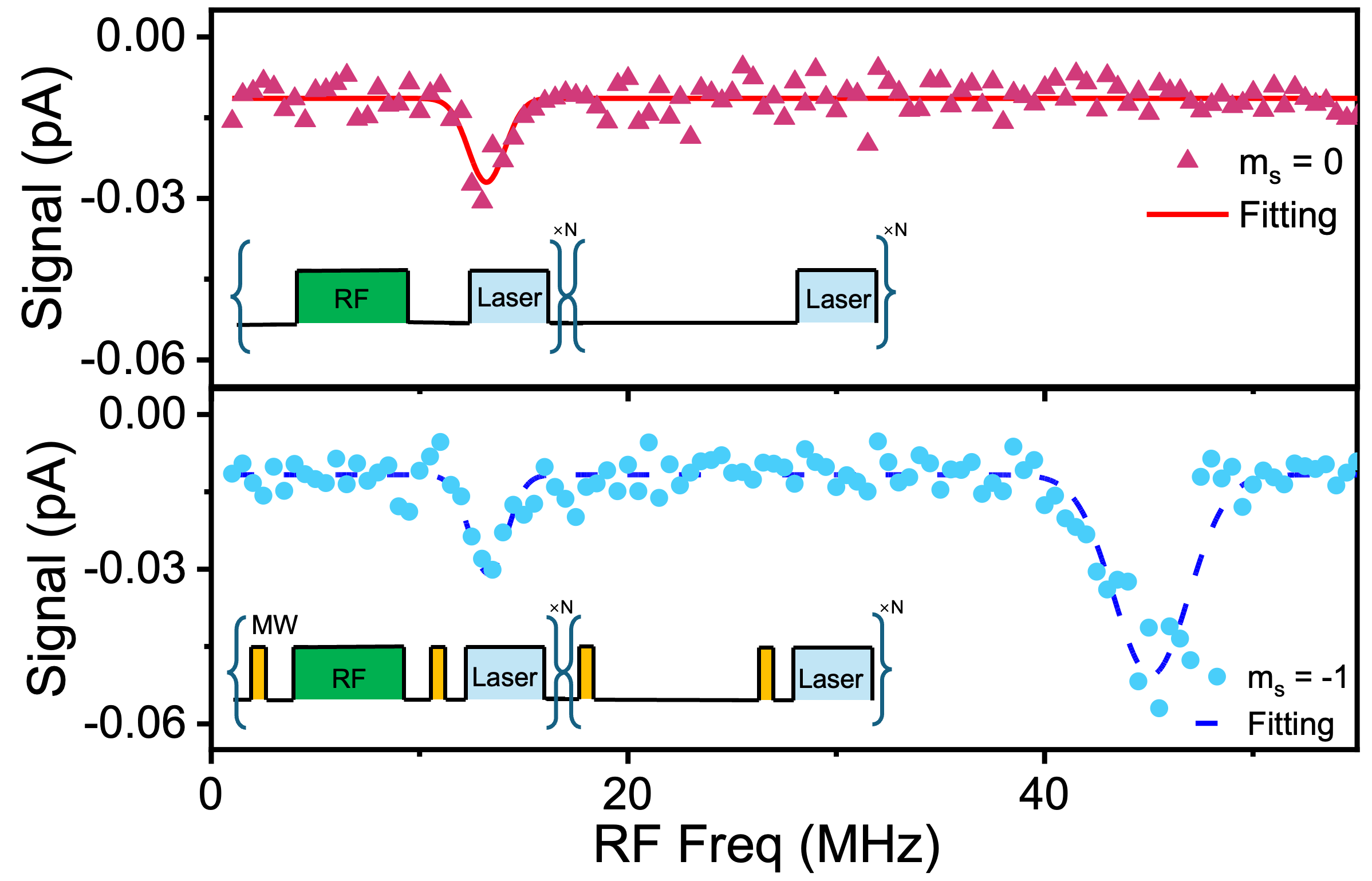}
  \caption{Photoelectric readout of nuclear spins. The upper and lower subpanels correspond to the $m_s = 0$ and $m_s = -1$ states, respectively, with timing sequences shown in the lower-left corner of each. For the $m_s = 0$ state, nuclear resonance is detected by sweeping the RF field. For the $m_s = -1$ state, an external MW $\pi$ pulse is applied to flip the electron spin before RF excitation. Red triangles represent data for the $m_s = 0$ state and blue circles for $m_s = -1$, with corresponding Gaussian fits shown as red and blue curves.}
  \label{fig: figure4}
\end{figure}
Nuclear spin polarization via two-quantum transitions enables direct readout of the nuclear spin state, as only specific configurations avoid energy level mixing and ISC-induced dark states, forming the basis for effective ENDOR detection \cite{RN37}.
Figure \ref{fig: figure4} displays, in each subpanel, the corresponding timing sequence for photoelectric nuclear spin readout in its lower-left corner. For $m_s=0$ state, the nuclear spin resonance frequency is determined by sweeping the RF frequency and comparing the photocurrent with and without the applied RF signal.
The resulting photoelectric signal is shown as red triangular markers in the upper panel of Fig. \ref{fig: figure4}. For $m_s=-1$ state, additional MW $\pi$ pulses are required to flip the electric spin and flop it back to its original state. The corresponding photoelectric ENDOR results are shown as blue circular markers in the lower panel of Fig. \ref{fig: figure4}. We then fit the data using Gaussian line shapes, extracting a hyperfine interaction of 13.2 MHz for the $m_s=0$ state and 45.1 MHz for the $m_s=-1$ state. Although efforts were made to place the wire antenna as close as possible to the measured device, the SNR of the nuclear spin signal remains significantly lower than that of the electronic spin signal. In future work, monolithic integration of the MW and RF antennas with the electrical readout electrodes on a single chip is expected to reduce crosstalk and significantly improve the SNR of the measurements.

\textit{Conclusion.}
In summary, we have demonstrated the PDMR of $\mathrm{V_B^-}$ defect spins in hBN. These findings underscore the promising potential of photoelectric detection for coherently driven boron vacancies in hBN. Unlike conventional optical techniques, this photoelectric approach eliminates the need for confocal detection and precise calibration, facilitating the development of compact magnetometers and electrometers.
For $\mathrm{V_B^-}$ defects, recent studies have shown that the nearby nitrogen nuclear spins can be polarized through excited- or ground-state energy level anti-crossings \cite{RN35, RN37, RN38, RN39}. Owing to their weak coupling with the surrounding environment, nuclear spins exhibit long coherence times, making them attractive candidates for quantum registers \cite{RN40, RN41}. By integrating photoelectric readout technology, it is possible to overcome the limitations of electron spin coherence time while enhancing the spatial resolution and sensitivity of spin-based quantum sensors \cite{RN42}. Moreover, the recent discovery of more spin-controllable single-photon sources operating at room temperature in hBN and other 2D materials further highlights the growing importance of this research area \cite{RN8, RN43, RN44}. Exploring in-plane rather than out-of-plane electrode geometries may allow the PDMR technique to achieve single-spin readout beyond the diffraction limit of optical spot, paving the way for higher-precision quantum sensing.

\begin{acknowledgments}
S.R., A.L., Z.J., Z.Li., X.L., Y.Y., J.Z.-P., and W.G. acknowledge support by ASTAR (M24M8b0004), Singapore National Research foundation (NRF-CRP22-2019-0004, NRF-CRP30-2023-0003, NRF-CRP31-0001, NRF2023-ITC004-001 and NRF-MSG-2023-0002) and Singapore Ministry of Education Tier 2 Grant (MOE-T2EP50222-0018).
F.J. acknowledges support by the German Federal Ministry of Research (BMBF) by future cluster QSENS (No. 03ZK110AB), European Union's HORIZON Europe program via projects QuMicro (No. 101046911), European Research Council (ERC) via Synergy grant HyperQ (No. 856432) and Carl-Zeiss-Stiftung via the Center of Integrated Quantum Science and technology (IQST).
H.L. and A.B. acknowledge support by Singapore Ministry of Education Tier 2 Grant (MOE-T2EP50221-0009).
\end{acknowledgments}


\begin{thebibliography}{45}%
\makeatletter
\providecommand \@ifxundefined [1]{%
 \@ifx{#1\undefined}
}%
\providecommand \@ifnum [1]{%
 \ifnum #1\expandafter \@firstoftwo
 \else \expandafter \@secondoftwo
 \fi
}%
\providecommand \@ifx [1]{%
 \ifx #1\expandafter \@firstoftwo
 \else \expandafter \@secondoftwo
 \fi
}%
\providecommand \natexlab [1]{#1}%
\providecommand \enquote  [1]{``#1''}%
\providecommand \bibnamefont  [1]{#1}%
\providecommand \bibfnamefont [1]{#1}%
\providecommand \citenamefont [1]{#1}%
\providecommand \href@noop [0]{\@secondoftwo}%
\providecommand \href [0]{\begingroup \@sanitize@url \@href}%
\providecommand \@href[1]{\@@startlink{#1}\@@href}%
\providecommand \@@href[1]{\endgroup#1\@@endlink}%
\providecommand \@sanitize@url [0]{\catcode `\\12\catcode `\$12\catcode
  `\&12\catcode `\#12\catcode `\^12\catcode `\_12\catcode `\%12\relax}%
\providecommand \@@startlink[1]{}%
\providecommand \@@endlink[0]{}%
\providecommand \url  [0]{\begingroup\@sanitize@url \@url }%
\providecommand \@url [1]{\endgroup\@href {#1}{\urlprefix }}%
\providecommand \urlprefix  [0]{URL }%
\providecommand \Eprint [0]{\href }%
\providecommand \doibase [0]{https://doi.org/}%
\providecommand \selectlanguage [0]{\@gobble}%
\providecommand \bibinfo  [0]{\@secondoftwo}%
\providecommand \bibfield  [0]{\@secondoftwo}%
\providecommand \translation [1]{[#1]}%
\providecommand \BibitemOpen [0]{}%
\providecommand \bibitemStop [0]{}%
\providecommand \bibitemNoStop [0]{.\EOS\space}%
\providecommand \EOS [0]{\spacefactor3000\relax}%
\providecommand \BibitemShut  [1]{\csname bibitem#1\endcsname}%
\let\auto@bib@innerbib\@empty
%</preamble>
\bibitem [{\citenamefont {Fang}\ \emph {et~al.}(2024)\citenamefont {Fang},
  \citenamefont {Wang}, \citenamefont {Marie},\ and\ \citenamefont
  {Sun}}]{RN1}%
  \BibitemOpen
  \bibfield  {author} {\bibinfo {author} {\bibfnamefont {H.-H.}\ \bibnamefont
  {Fang}}, \bibinfo {author} {\bibfnamefont {X.-J.}\ \bibnamefont {Wang}},
  \bibinfo {author} {\bibfnamefont {X.}~\bibnamefont {Marie}},\ and\ \bibinfo
  {author} {\bibfnamefont {H.-B.}\ \bibnamefont {Sun}},\ }\bibfield  {title}
  {\bibinfo {title} {Quantum sensing with optically accessible spin defects in
  van der waals layered materials},\ }\href
  {https://doi.org/10.1038/s41377-024-01630-y} {\bibfield  {journal} {\bibinfo
  {journal} {Light: Science \& Applications}\ }\textbf {\bibinfo {volume}
  {13}},\ \bibinfo {pages} {303} (\bibinfo {year} {2024})}\BibitemShut
  {NoStop}%
\bibitem [{\citenamefont {Turunen}\ \emph {et~al.}(2022)\citenamefont
  {Turunen}, \citenamefont {Brotons-Gisbert}, \citenamefont {Dai},
  \citenamefont {Wang}, \citenamefont {Scerri}, \citenamefont {Bonato},
  \citenamefont {J\"ons}, \citenamefont {Sun},\ and\ \citenamefont
  {Gerardot}}]{RN2}%
  \BibitemOpen
  \bibfield  {author} {\bibinfo {author} {\bibfnamefont {M.}~\bibnamefont
  {Turunen}}, \bibinfo {author} {\bibfnamefont {M.}~\bibnamefont
  {Brotons-Gisbert}}, \bibinfo {author} {\bibfnamefont {Y.}~\bibnamefont
  {Dai}}, \bibinfo {author} {\bibfnamefont {Y.}~\bibnamefont {Wang}}, \bibinfo
  {author} {\bibfnamefont {E.}~\bibnamefont {Scerri}}, \bibinfo {author}
  {\bibfnamefont {C.}~\bibnamefont {Bonato}}, \bibinfo {author} {\bibfnamefont
  {K.~D.}\ \bibnamefont {J\"ons}}, \bibinfo {author} {\bibfnamefont
  {Z.}~\bibnamefont {Sun}},\ and\ \bibinfo {author} {\bibfnamefont {B.~D.}\
  \bibnamefont {Gerardot}},\ }\bibfield  {title} {\bibinfo {title} {Quantum
  photonics with layered 2D materials},\ }\href
  {https://doi.org/10.1038/s42254-021-00408-0} {\bibfield  {journal} {\bibinfo
  {journal} {Nature Reviews Physics}\ }\textbf {\bibinfo {volume} {4}},\
  \bibinfo {pages} {219} (\bibinfo {year} {2022})}\BibitemShut {NoStop}%
\bibitem [{\citenamefont {Liu}\ and\ \citenamefont {Hersam}(2019)}]{RN3}%
  \BibitemOpen
  \bibfield  {author} {\bibinfo {author} {\bibfnamefont {X.}~\bibnamefont
  {Liu}}\ and\ \bibinfo {author} {\bibfnamefont {M.~C.}\ \bibnamefont
  {Hersam}},\ }\bibfield  {title} {\bibinfo {title} {2D materials for quantum
  information science},\ }\href {https://doi.org/10.1038/s41578-019-0136-x}
  {\bibfield  {journal} {\bibinfo  {journal} {Nature Reviews Materials}\
  }\textbf {\bibinfo {volume} {4}},\ \bibinfo {pages} {669} (\bibinfo {year}
  {2019})}\BibitemShut {NoStop}%
\bibitem [{\citenamefont {Brenneis}\ \emph {et~al.}(2015)\citenamefont
  {Brenneis}, \citenamefont {Gaudreau}, \citenamefont {Seifert}, \citenamefont
  {Karl}, \citenamefont {Brandt}, \citenamefont {Huebl}, \citenamefont
  {Garrido}, \citenamefont {Koppens},\ and\ \citenamefont {Holleitner}}]{RN4}%
  \BibitemOpen
  \bibfield  {author} {\bibinfo {author} {\bibfnamefont {A.}~\bibnamefont
  {Brenneis}}, \bibinfo {author} {\bibfnamefont {L.}~\bibnamefont {Gaudreau}},
  \bibinfo {author} {\bibfnamefont {M.}~\bibnamefont {Seifert}}, \bibinfo
  {author} {\bibfnamefont {H.}~\bibnamefont {Karl}}, \bibinfo {author}
  {\bibfnamefont {M.~S.}\ \bibnamefont {Brandt}}, \bibinfo {author}
  {\bibfnamefont {H.}~\bibnamefont {Huebl}}, \bibinfo {author} {\bibfnamefont
  {J.~A.}\ \bibnamefont {Garrido}}, \bibinfo {author} {\bibfnamefont
  {F.~H.~L.}\ \bibnamefont {Koppens}},\ and\ \bibinfo {author} {\bibfnamefont
  {A.~W.}\ \bibnamefont {Holleitner}},\ }\bibfield  {title} {\bibinfo {title}
  {Ultrafast electronic readout of diamond Nitrogen–Vacancy centres coupled
  to graphene},\ }\href {https://doi.org/10.1038/nnano.2014.276} {\bibfield
  {journal} {\bibinfo  {journal} {Nature Nanotechnology}\ }\textbf {\bibinfo
  {volume} {10}},\ \bibinfo {pages} {135} (\bibinfo {year} {2015})}\BibitemShut
  {NoStop}%
\bibitem [{\citenamefont {Liu}\ \emph {et~al.}(2022)\citenamefont {Liu},
  \citenamefont {Guo}, \citenamefont {Yu}, \citenamefont {Meng}, \citenamefont
  {Li}, \citenamefont {Yang}, \citenamefont {Wang}, \citenamefont {Zeng},
  \citenamefont {Xie}, \citenamefont {Li}, \citenamefont {Wang}, \citenamefont
  {Xu}, \citenamefont {Wang}, \citenamefont {Tang}, \citenamefont {Li},\ and\
  \citenamefont {Guo}}]{RN5}%
  \BibitemOpen
  \bibfield  {author} {\bibinfo {author} {\bibfnamefont {W.}~\bibnamefont
  {Liu}}, \bibinfo {author} {\bibfnamefont {N.-J.}\ \bibnamefont {Guo}},
  \bibinfo {author} {\bibfnamefont {S.}~\bibnamefont {Yu}}, \bibinfo {author}
  {\bibfnamefont {Y.}~\bibnamefont {Meng}}, \bibinfo {author} {\bibfnamefont
  {Z.-P.}\ \bibnamefont {Li}}, \bibinfo {author} {\bibfnamefont {Y.-Z.}\
  \bibnamefont {Yang}}, \bibinfo {author} {\bibfnamefont {Z.-A.}\ \bibnamefont
  {Wang}}, \bibinfo {author} {\bibfnamefont {X.-D.}\ \bibnamefont {Zeng}},
  \bibinfo {author} {\bibfnamefont {L.-K.}\ \bibnamefont {Xie}}, \bibinfo
  {author} {\bibfnamefont {Q.}~\bibnamefont {Li}}, \bibinfo {author}
  {\bibfnamefont {J.-F.}\ \bibnamefont {Wang}}, \bibinfo {author}
  {\bibfnamefont {J.-S.}\ \bibnamefont {Xu}}, \bibinfo {author} {\bibfnamefont
  {Y.-T.}\ \bibnamefont {Wang}}, \bibinfo {author} {\bibfnamefont {J.-S.}\
  \bibnamefont {Tang}}, \bibinfo {author} {\bibfnamefont {C.-F.}\ \bibnamefont
  {Li}},\ and\ \bibinfo {author} {\bibfnamefont {G.-C.}\ \bibnamefont {Guo}},\
  }\bibfield  {title} {\bibinfo {title} {Spin-active defects in hexagonal Boron
  Nitride},\ }\href {https://doi.org/10.1088/2633-4356/ac7e9f} {\bibfield
  {journal} {\bibinfo  {journal} {Materials for Quantum Technology}\ }\textbf
  {\bibinfo {volume} {2}},\ \bibinfo {pages} {032002} (\bibinfo {year}
  {2022})}\BibitemShut {NoStop}%
\bibitem [{\citenamefont {Gottscholl}\ \emph
  {et~al.}(2021{\natexlab{a}})\citenamefont {Gottscholl}, \citenamefont {Diez},
  \citenamefont {Soltamov}, \citenamefont {Kasper}, \citenamefont {Sperlich},
  \citenamefont {Kianinia}, \citenamefont {Bradac}, \citenamefont
  {Aharonovich},\ and\ \citenamefont {Dyakonov}}]{RN6}%
  \BibitemOpen
  \bibfield  {author} {\bibinfo {author} {\bibfnamefont {A.}~\bibnamefont
  {Gottscholl}}, \bibinfo {author} {\bibfnamefont {M.}~\bibnamefont {Diez}},
  \bibinfo {author} {\bibfnamefont {V.}~\bibnamefont {Soltamov}}, \bibinfo
  {author} {\bibfnamefont {C.}~\bibnamefont {Kasper}}, \bibinfo {author}
  {\bibfnamefont {A.}~\bibnamefont {Sperlich}}, \bibinfo {author}
  {\bibfnamefont {M.}~\bibnamefont {Kianinia}}, \bibinfo {author}
  {\bibfnamefont {C.}~\bibnamefont {Bradac}}, \bibinfo {author} {\bibfnamefont
  {I.}~\bibnamefont {Aharonovich}},\ and\ \bibinfo {author} {\bibfnamefont
  {V.}~\bibnamefont {Dyakonov}},\ }\bibfield  {title} {\bibinfo {title} {Room
  temperature coherent control of spin defects in hexagonal boron nitride},\
  }\href {https://doi.org/10.1126/sciadv.abf3630} {\bibfield  {journal}
  {\bibinfo  {journal} {Science Advances}\ }\textbf {\bibinfo {volume} {7}},\
  \bibinfo {pages} {eabf3630} (\bibinfo {year}
  {2021}{\natexlab{a}})}\BibitemShut {NoStop}%
\bibitem [{\citenamefont {Gottscholl}\ \emph {et~al.}(2020)\citenamefont
  {Gottscholl}, \citenamefont {Kianinia}, \citenamefont {Soltamov},
  \citenamefont {Orlinskii}, \citenamefont {Mamin}, \citenamefont {Bradac},
  \citenamefont {Kasper}, \citenamefont {Krambrock}, \citenamefont {Sperlich},
  \citenamefont {Toth}, \citenamefont {Aharonovich},\ and\ \citenamefont
  {Dyakonov}}]{RN7}%
  \BibitemOpen
  \bibfield  {author} {\bibinfo {author} {\bibfnamefont {A.}~\bibnamefont
  {Gottscholl}}, \bibinfo {author} {\bibfnamefont {M.}~\bibnamefont
  {Kianinia}}, \bibinfo {author} {\bibfnamefont {V.}~\bibnamefont {Soltamov}},
  \bibinfo {author} {\bibfnamefont {S.}~\bibnamefont {Orlinskii}}, \bibinfo
  {author} {\bibfnamefont {G.}~\bibnamefont {Mamin}}, \bibinfo {author}
  {\bibfnamefont {C.}~\bibnamefont {Bradac}}, \bibinfo {author} {\bibfnamefont
  {C.}~\bibnamefont {Kasper}}, \bibinfo {author} {\bibfnamefont
  {K.}~\bibnamefont {Krambrock}}, \bibinfo {author} {\bibfnamefont
  {A.}~\bibnamefont {Sperlich}}, \bibinfo {author} {\bibfnamefont
  {M.}~\bibnamefont {Toth}}, \bibinfo {author} {\bibfnamefont {I.}~\bibnamefont
  {Aharonovich}},\ and\ \bibinfo {author} {\bibfnamefont {V.}~\bibnamefont
  {Dyakonov}},\ }\bibfield  {title} {\bibinfo {title} {Initialization and
  read-out of intrinsic spin defects in a van der waals crystal at room
  temperature},\ }\href {https://doi.org/10.1038/s41563-020-0619-6} {\bibfield
  {journal} {\bibinfo  {journal} {Nature Materials}\ }\textbf {\bibinfo
  {volume} {19}},\ \bibinfo {pages} {540} (\bibinfo {year} {2020})}\BibitemShut
  {NoStop}%
\bibitem [{\citenamefont {Stern}\ \emph {et~al.}(2024)\citenamefont {Stern},
  \citenamefont {M.~Gilardoni}, \citenamefont {Gu}, \citenamefont
  {Eizagirre~Barker}, \citenamefont {Powell}, \citenamefont {Deng},
  \citenamefont {Fraser}, \citenamefont {Follet}, \citenamefont {Li},
  \citenamefont {Ramsay}, \citenamefont {Tan}, \citenamefont {Aharonovich},\
  and\ \citenamefont {Atatüre}}]{RN8}%
  \BibitemOpen
  \bibfield  {author} {\bibinfo {author} {\bibfnamefont {H.~L.}\ \bibnamefont
  {Stern}}, \bibinfo {author} {\bibfnamefont {C.}~\bibnamefont {M.~Gilardoni}},
  \bibinfo {author} {\bibfnamefont {Q.}~\bibnamefont {Gu}}, \bibinfo {author}
  {\bibfnamefont {S.}~\bibnamefont {Eizagirre~Barker}}, \bibinfo {author}
  {\bibfnamefont {O.~F.~J.}\ \bibnamefont {Powell}}, \bibinfo {author}
  {\bibfnamefont {X.}~\bibnamefont {Deng}}, \bibinfo {author} {\bibfnamefont
  {S.~A.}\ \bibnamefont {Fraser}}, \bibinfo {author} {\bibfnamefont
  {L.}~\bibnamefont {Follet}}, \bibinfo {author} {\bibfnamefont
  {C.}~\bibnamefont {Li}}, \bibinfo {author} {\bibfnamefont {A.~J.}\
  \bibnamefont {Ramsay}}, \bibinfo {author} {\bibfnamefont {H.~H.}\
  \bibnamefont {Tan}}, \bibinfo {author} {\bibfnamefont {I.}~\bibnamefont
  {Aharonovich}},\ and\ \bibinfo {author} {\bibfnamefont {M.}~\bibnamefont
  {Atatüre}},\ }\bibfield  {title} {\bibinfo {title} {A quantum coherent spin
  in hexagonal boron nitride at ambient conditions},\ }\href
  {https://doi.org/10.1038/s41563-024-01887-z} {\bibfield  {journal} {\bibinfo
  {journal} {Nature Materials}\ }\textbf {\bibinfo {volume} {23}},\ \bibinfo
  {pages} {1379} (\bibinfo {year} {2024})}\BibitemShut {NoStop}%
\bibitem [{\citenamefont {Zhou}\ \emph {et~al.}(2023)\citenamefont {Zhou},
  \citenamefont {Jiang}, \citenamefont {Liang}, \citenamefont {Ru},
  \citenamefont {Bettiol},\ and\ \citenamefont {Gao}}]{RN9}%
  \BibitemOpen
  \bibfield  {author} {\bibinfo {author} {\bibfnamefont {F.}~\bibnamefont
  {Zhou}}, \bibinfo {author} {\bibfnamefont {Z.}~\bibnamefont {Jiang}},
  \bibinfo {author} {\bibfnamefont {H.}~\bibnamefont {Liang}}, \bibinfo
  {author} {\bibfnamefont {S.}~\bibnamefont {Ru}}, \bibinfo {author}
  {\bibfnamefont {A.~A.}\ \bibnamefont {Bettiol}},\ and\ \bibinfo {author}
  {\bibfnamefont {W.}~\bibnamefont {Gao}},\ }\bibfield  {title} {\bibinfo
  {title} {Dc magnetic field sensitivity optimization of spin defects in
  hexagonal boron nitride},\ }\href
  {https://doi.org/10.1021/acs.nanolett.3c01881} {\bibfield  {journal}
  {\bibinfo  {journal} {Nano Letters}\ }\textbf {\bibinfo {volume} {23}},\
  \bibinfo {pages} {6209} (\bibinfo {year} {2023})}\BibitemShut {NoStop}%
\bibitem [{\citenamefont {Gottscholl}\ \emph
  {et~al.}(2021{\natexlab{b}})\citenamefont {Gottscholl}, \citenamefont {Diez},
  \citenamefont {Soltamov}, \citenamefont {Kasper}, \citenamefont {Krauße},
  \citenamefont {Sperlich}, \citenamefont {Kianinia}, \citenamefont {Bradac},
  \citenamefont {Aharonovich},\ and\ \citenamefont {Dyakonov}}]{RN10}%
  \BibitemOpen
  \bibfield  {author} {\bibinfo {author} {\bibfnamefont {A.}~\bibnamefont
  {Gottscholl}}, \bibinfo {author} {\bibfnamefont {M.}~\bibnamefont {Diez}},
  \bibinfo {author} {\bibfnamefont {V.}~\bibnamefont {Soltamov}}, \bibinfo
  {author} {\bibfnamefont {C.}~\bibnamefont {Kasper}}, \bibinfo {author}
  {\bibfnamefont {D.}~\bibnamefont {Krauße}}, \bibinfo {author} {\bibfnamefont
  {A.}~\bibnamefont {Sperlich}}, \bibinfo {author} {\bibfnamefont
  {M.}~\bibnamefont {Kianinia}}, \bibinfo {author} {\bibfnamefont
  {C.}~\bibnamefont {Bradac}}, \bibinfo {author} {\bibfnamefont
  {I.}~\bibnamefont {Aharonovich}},\ and\ \bibinfo {author} {\bibfnamefont
  {V.}~\bibnamefont {Dyakonov}},\ }\bibfield  {title} {\bibinfo {title} {Spin
  defects in hbn as promising temperature, pressure and magnetic field quantum
  sensors},\ }\href {https://doi.org/10.1038/s41467-021-24725-1} {\bibfield
  {journal} {\bibinfo  {journal} {Nature Communications}\ }\textbf {\bibinfo
  {volume} {12}},\ \bibinfo {pages} {4480} (\bibinfo {year}
  {2021}{\natexlab{b}})}\BibitemShut {NoStop}%
\bibitem [{\citenamefont {Lyu}\ \emph {et~al.}(2022)\citenamefont {Lyu},
  \citenamefont {Tan}, \citenamefont {Wu}, \citenamefont {Zhang}, \citenamefont
  {Zhang}, \citenamefont {Mu}, \citenamefont {Zúñiga-Pérez}, \citenamefont
  {Cai},\ and\ \citenamefont {Gao}}]{RN11}%
  \BibitemOpen
  \bibfield  {author} {\bibinfo {author} {\bibfnamefont {X.}~\bibnamefont
  {Lyu}}, \bibinfo {author} {\bibfnamefont {Q.}~\bibnamefont {Tan}}, \bibinfo
  {author} {\bibfnamefont {L.}~\bibnamefont {Wu}}, \bibinfo {author}
  {\bibfnamefont {C.}~\bibnamefont {Zhang}}, \bibinfo {author} {\bibfnamefont
  {Z.}~\bibnamefont {Zhang}}, \bibinfo {author} {\bibfnamefont
  {Z.}~\bibnamefont {Mu}}, \bibinfo {author} {\bibfnamefont {J.}~\bibnamefont
  {Zúñiga-Pérez}}, \bibinfo {author} {\bibfnamefont {H.}~\bibnamefont
  {Cai}},\ and\ \bibinfo {author} {\bibfnamefont {W.}~\bibnamefont {Gao}},\
  }\bibfield  {title} {\bibinfo {title} {Strain quantum sensing with spin
  defects in hexagonal boron nitride},\ }\href
  {https://doi.org/10.1021/acs.nanolett.2c01722} {\bibfield  {journal}
  {\bibinfo  {journal} {Nano Letters}\ }\textbf {\bibinfo {volume} {22}},\
  \bibinfo {pages} {6553} (\bibinfo {year} {2022})}\BibitemShut {NoStop}%
\bibitem [{\citenamefont {Healey}\ \emph {et~al.}(2023)\citenamefont {Healey},
  \citenamefont {Scholten}, \citenamefont {Yang}, \citenamefont {Scott},
  \citenamefont {Abrahams}, \citenamefont {Robertson}, \citenamefont {Hou},
  \citenamefont {Guo}, \citenamefont {Rahman}, \citenamefont {Lu},
  \citenamefont {Kianinia}, \citenamefont {Aharonovich},\ and\ \citenamefont
  {Tetienne}}]{RN12}%
  \BibitemOpen
  \bibfield  {author} {\bibinfo {author} {\bibfnamefont {A.~J.}\ \bibnamefont
  {Healey}}, \bibinfo {author} {\bibfnamefont {S.~C.}\ \bibnamefont
  {Scholten}}, \bibinfo {author} {\bibfnamefont {T.}~\bibnamefont {Yang}},
  \bibinfo {author} {\bibfnamefont {J.~A.}\ \bibnamefont {Scott}}, \bibinfo
  {author} {\bibfnamefont {G.~J.}\ \bibnamefont {Abrahams}}, \bibinfo {author}
  {\bibfnamefont {I.~O.}\ \bibnamefont {Robertson}}, \bibinfo {author}
  {\bibfnamefont {X.~F.}\ \bibnamefont {Hou}}, \bibinfo {author} {\bibfnamefont
  {Y.~F.}\ \bibnamefont {Guo}}, \bibinfo {author} {\bibfnamefont
  {S.}~\bibnamefont {Rahman}}, \bibinfo {author} {\bibfnamefont
  {Y.}~\bibnamefont {Lu}}, \bibinfo {author} {\bibfnamefont {M.}~\bibnamefont
  {Kianinia}}, \bibinfo {author} {\bibfnamefont {I.}~\bibnamefont
  {Aharonovich}},\ and\ \bibinfo {author} {\bibfnamefont {J.~P.}\ \bibnamefont
  {Tetienne}},\ }\bibfield  {title} {\bibinfo {title} {Quantum microscopy with
  van der waals heterostructures},\ }\href
  {https://doi.org/10.1038/s41567-022-01815-5} {\bibfield  {journal} {\bibinfo
  {journal} {Nature Physics}\ }\textbf {\bibinfo {volume} {19}},\ \bibinfo
  {pages} {87} (\bibinfo {year} {2023})}\BibitemShut {NoStop}%
\bibitem [{\citenamefont {Liu}\ \emph {et~al.}(2021)\citenamefont {Liu},
  \citenamefont {Li}, \citenamefont {Yang}, \citenamefont {Yu}, \citenamefont
  {Meng}, \citenamefont {Wang}, \citenamefont {Li}, \citenamefont {Guo},
  \citenamefont {Yan}, \citenamefont {Li}, \citenamefont {Wang}, \citenamefont
  {Xu}, \citenamefont {Wang}, \citenamefont {Tang}, \citenamefont {Li},\ and\
  \citenamefont {Guo}}]{RN13}%
  \BibitemOpen
  \bibfield  {author} {\bibinfo {author} {\bibfnamefont {W.}~\bibnamefont
  {Liu}}, \bibinfo {author} {\bibfnamefont {Z.-P.}\ \bibnamefont {Li}},
  \bibinfo {author} {\bibfnamefont {Y.-Z.}\ \bibnamefont {Yang}}, \bibinfo
  {author} {\bibfnamefont {S.}~\bibnamefont {Yu}}, \bibinfo {author}
  {\bibfnamefont {Y.}~\bibnamefont {Meng}}, \bibinfo {author} {\bibfnamefont
  {Z.-A.}\ \bibnamefont {Wang}}, \bibinfo {author} {\bibfnamefont {Z.-C.}\
  \bibnamefont {Li}}, \bibinfo {author} {\bibfnamefont {N.-J.}\ \bibnamefont
  {Guo}}, \bibinfo {author} {\bibfnamefont {F.-F.}\ \bibnamefont {Yan}},
  \bibinfo {author} {\bibfnamefont {Q.}~\bibnamefont {Li}}, \bibinfo {author}
  {\bibfnamefont {J.-F.}\ \bibnamefont {Wang}}, \bibinfo {author}
  {\bibfnamefont {J.-S.}\ \bibnamefont {Xu}}, \bibinfo {author} {\bibfnamefont
  {Y.-T.}\ \bibnamefont {Wang}}, \bibinfo {author} {\bibfnamefont {J.-S.}\
  \bibnamefont {Tang}}, \bibinfo {author} {\bibfnamefont {C.-F.}\ \bibnamefont
  {Li}},\ and\ \bibinfo {author} {\bibfnamefont {G.-C.}\ \bibnamefont {Guo}},\
  }\bibfield  {title} {\bibinfo {title} {Temperature-dependent energy-level
  shifts of spin defects in hexagonal boron nitride},\ }\href
  {https://doi.org/10.1021/acsphotonics.1c00320} {\bibfield  {journal}
  {\bibinfo  {journal} {ACS Photonics}\ }\textbf {\bibinfo {volume} {8}},\
  \bibinfo {pages} {1889} (\bibinfo {year} {2021})}\BibitemShut {NoStop}%
\bibitem [{\citenamefont {Vaidya}\ \emph {et~al.}(2023)\citenamefont {Vaidya},
  \citenamefont {Gao}, \citenamefont {Dikshit}, \citenamefont {Aharonovich},\
  and\ \citenamefont {Li}}]{RN14}%
  \BibitemOpen
  \bibfield  {author} {\bibinfo {author} {\bibfnamefont {S.}~\bibnamefont
  {Vaidya}}, \bibinfo {author} {\bibfnamefont {X.}~\bibnamefont {Gao}},
  \bibinfo {author} {\bibfnamefont {S.}~\bibnamefont {Dikshit}}, \bibinfo
  {author} {\bibfnamefont {I.}~\bibnamefont {Aharonovich}},\ and\ \bibinfo
  {author} {\bibfnamefont {T.}~\bibnamefont {Li}},\ }\bibfield  {title}
  {\bibinfo {title} {Quantum sensing and imaging with spin defects in hexagonal
  boron nitride},\ }\href {https://doi.org/10.1080/23746149.2023.2206049}
  {\bibfield  {journal} {\bibinfo  {journal} {Advances in Physics: X}\ }\textbf
  {\bibinfo {volume} {8}},\ \bibinfo {pages} {2206049} (\bibinfo {year}
  {2023})}\BibitemShut {NoStop}%
\bibitem [{\citenamefont {Caldwell}\ \emph {et~al.}(2019)\citenamefont
  {Caldwell}, \citenamefont {Aharonovich}, \citenamefont {Cassabois},
  \citenamefont {Edgar}, \citenamefont {Gil},\ and\ \citenamefont
  {Basov}}]{RN15}%
  \BibitemOpen
  \bibfield  {author} {\bibinfo {author} {\bibfnamefont {J.~D.}\ \bibnamefont
  {Caldwell}}, \bibinfo {author} {\bibfnamefont {I.}~\bibnamefont
  {Aharonovich}}, \bibinfo {author} {\bibfnamefont {G.}~\bibnamefont
  {Cassabois}}, \bibinfo {author} {\bibfnamefont {J.~H.}\ \bibnamefont
  {Edgar}}, \bibinfo {author} {\bibfnamefont {B.}~\bibnamefont {Gil}},\ and\
  \bibinfo {author} {\bibfnamefont {D.~N.}\ \bibnamefont {Basov}},\ }\bibfield
  {title} {\bibinfo {title} {Photonics with hexagonal boron nitride},\ }\href
  {https://doi.org/10.1038/s41578-019-0124-1} {\bibfield  {journal} {\bibinfo
  {journal} {Nature Reviews Materials}\ }\textbf {\bibinfo {volume} {4}},\
  \bibinfo {pages} {552} (\bibinfo {year} {2019})}\BibitemShut {NoStop}%
\bibitem [{\citenamefont {Gao}\ \emph {et~al.}(2024)\citenamefont {Gao},
  \citenamefont {Vaidya}, \citenamefont {Dikshit}, \citenamefont {Ju},
  \citenamefont {Shen}, \citenamefont {Jin}, \citenamefont {Zhang},\ and\
  \citenamefont {Li}}]{RN16}%
  \BibitemOpen
  \bibfield  {author} {\bibinfo {author} {\bibfnamefont {X.}~\bibnamefont
  {Gao}}, \bibinfo {author} {\bibfnamefont {S.}~\bibnamefont {Vaidya}},
  \bibinfo {author} {\bibfnamefont {S.}~\bibnamefont {Dikshit}}, \bibinfo
  {author} {\bibfnamefont {P.}~\bibnamefont {Ju}}, \bibinfo {author}
  {\bibfnamefont {K.}~\bibnamefont {Shen}}, \bibinfo {author} {\bibfnamefont
  {Y.}~\bibnamefont {Jin}}, \bibinfo {author} {\bibfnamefont {S.}~\bibnamefont
  {Zhang}},\ and\ \bibinfo {author} {\bibfnamefont {T.}~\bibnamefont {Li}},\
  }\bibfield  {title} {\bibinfo {title} {Nanotube spin defects for
  omnidirectional magnetic field sensing},\ }\href
  {https://doi.org/10.1038/s41467-024-51941-2} {\bibfield  {journal} {\bibinfo
  {journal} {Nature Communications}\ }\textbf {\bibinfo {volume} {15}},\
  \bibinfo {pages} {7697} (\bibinfo {year} {2024})}\BibitemShut {NoStop}%
\bibitem [{\citenamefont {Huang}\ \emph {et~al.}(2022)\citenamefont {Huang},
  \citenamefont {Zhou}, \citenamefont {Chen}, \citenamefont {Lu}, \citenamefont
  {McLaughlin}, \citenamefont {Li}, \citenamefont {Alghamdi}, \citenamefont
  {Djugba}, \citenamefont {Shi}, \citenamefont {Wang},\ and\ \citenamefont
  {Du}}]{RN17}%
  \BibitemOpen
  \bibfield  {author} {\bibinfo {author} {\bibfnamefont {M.}~\bibnamefont
  {Huang}}, \bibinfo {author} {\bibfnamefont {J.}~\bibnamefont {Zhou}},
  \bibinfo {author} {\bibfnamefont {D.}~\bibnamefont {Chen}}, \bibinfo {author}
  {\bibfnamefont {H.}~\bibnamefont {Lu}}, \bibinfo {author} {\bibfnamefont
  {N.~J.}\ \bibnamefont {McLaughlin}}, \bibinfo {author} {\bibfnamefont
  {S.}~\bibnamefont {Li}}, \bibinfo {author} {\bibfnamefont {M.}~\bibnamefont
  {Alghamdi}}, \bibinfo {author} {\bibfnamefont {D.}~\bibnamefont {Djugba}},
  \bibinfo {author} {\bibfnamefont {J.}~\bibnamefont {Shi}}, \bibinfo {author}
  {\bibfnamefont {H.}~\bibnamefont {Wang}},\ and\ \bibinfo {author}
  {\bibfnamefont {C.~R.}\ \bibnamefont {Du}},\ }\bibfield  {title} {\bibinfo
  {title} {Wide field imaging of van der waals ferromagnet fe3gete2 by spin
  defects in hexagonal boron nitride},\ }\href
  {https://doi.org/10.1038/s41467-022-33016-2} {\bibfield  {journal} {\bibinfo
  {journal} {Nature Communications}\ }\textbf {\bibinfo {volume} {13}},\
  \bibinfo {pages} {5369} (\bibinfo {year} {2022})}\BibitemShut {NoStop}%
\bibitem [{\citenamefont {Zhou}\ \emph {et~al.}(2024)\citenamefont {Zhou},
  \citenamefont {Lu}, \citenamefont {Chen}, \citenamefont {Huang},
  \citenamefont {Yan}, \citenamefont {Al-matouq}, \citenamefont {Chang},
  \citenamefont {Djugba}, \citenamefont {Jiang}, \citenamefont {Wang},\ and\
  \citenamefont {Du}}]{RN18}%
  \BibitemOpen
  \bibfield  {author} {\bibinfo {author} {\bibfnamefont {J.}~\bibnamefont
  {Zhou}}, \bibinfo {author} {\bibfnamefont {H.}~\bibnamefont {Lu}}, \bibinfo
  {author} {\bibfnamefont {D.}~\bibnamefont {Chen}}, \bibinfo {author}
  {\bibfnamefont {M.}~\bibnamefont {Huang}}, \bibinfo {author} {\bibfnamefont
  {G.~Q.}\ \bibnamefont {Yan}}, \bibinfo {author} {\bibfnamefont
  {F.}~\bibnamefont {Al-matouq}}, \bibinfo {author} {\bibfnamefont
  {J.}~\bibnamefont {Chang}}, \bibinfo {author} {\bibfnamefont
  {D.}~\bibnamefont {Djugba}}, \bibinfo {author} {\bibfnamefont
  {Z.}~\bibnamefont {Jiang}}, \bibinfo {author} {\bibfnamefont
  {H.}~\bibnamefont {Wang}},\ and\ \bibinfo {author} {\bibfnamefont {C.~R.}\
  \bibnamefont {Du}},\ }\bibfield  {title} {\bibinfo {title} {Sensing spin wave
  excitations by spin defects in few-layer-thick hexagonal boron nitride},\
  }\href {https://doi.org/10.1126/sciadv.adk8495} {\bibfield  {journal}
  {\bibinfo  {journal} {Science Advances}\ }\textbf {\bibinfo {volume} {10}},\
  \bibinfo {pages} {eadk8495} (\bibinfo {year} {2024})}\BibitemShut {NoStop}%
\bibitem [{\citenamefont {Durand}\ \emph {et~al.}(2023)\citenamefont {Durand},
  \citenamefont {Clua-Provost}, \citenamefont {Fabre}, \citenamefont {Kumar},
  \citenamefont {Li}, \citenamefont {Edgar}, \citenamefont {Udvarhelyi},
  \citenamefont {Gali}, \citenamefont {Marie}, \citenamefont {Robert},
  \citenamefont {G\'erard}, \citenamefont {Gil}, \citenamefont {Cassabois},\
  and\ \citenamefont {Jacques}}]{RN19}%
  \BibitemOpen
  \bibfield  {author} {\bibinfo {author} {\bibfnamefont {A.}~\bibnamefont
  {Durand}}, \bibinfo {author} {\bibfnamefont {T.}~\bibnamefont
  {Clua-Provost}}, \bibinfo {author} {\bibfnamefont {F.}~\bibnamefont {Fabre}},
  \bibinfo {author} {\bibfnamefont {P.}~\bibnamefont {Kumar}}, \bibinfo
  {author} {\bibfnamefont {J.}~\bibnamefont {Li}}, \bibinfo {author}
  {\bibfnamefont {J.~H.}\ \bibnamefont {Edgar}}, \bibinfo {author}
  {\bibfnamefont {P.}~\bibnamefont {Udvarhelyi}}, \bibinfo {author}
  {\bibfnamefont {A.}~\bibnamefont {Gali}}, \bibinfo {author} {\bibfnamefont
  {X.}~\bibnamefont {Marie}}, \bibinfo {author} {\bibfnamefont
  {C.}~\bibnamefont {Robert}}, \bibinfo {author} {\bibfnamefont {J.~M.}\
  \bibnamefont {G\'erard}}, \bibinfo {author} {\bibfnamefont {B.}~\bibnamefont
  {Gil}}, \bibinfo {author} {\bibfnamefont {G.}~\bibnamefont {Cassabois}},\
  and\ \bibinfo {author} {\bibfnamefont {V.}~\bibnamefont {Jacques}},\
  }\bibfield  {title} {\bibinfo {title} {Optically active spin defects in
  few-layer thick hexagonal boron nitride},\ }\href
  {https://doi.org/10.1103/PhysRevLett.131.116902} {\bibfield  {journal}
  {\bibinfo  {journal} {Physical Review Letters}\ }\textbf {\bibinfo {volume}
  {131}},\ \bibinfo {pages} {116902} (\bibinfo {year} {2023})}\BibitemShut
  {NoStop}%
\bibitem [{\citenamefont {Klein}\ \emph {et~al.}(2024)\citenamefont {Klein},
  \citenamefont {Zondiner}, \citenamefont {Keren}, \citenamefont {Birkbeck},
  \citenamefont {Inbar}, \citenamefont {Xiao}, \citenamefont {Sidorova},
  \citenamefont {Ezzi}, \citenamefont {Peng}, \citenamefont {Watanabe},
  \citenamefont {Taniguchi}, \citenamefont {Adam},\ and\ \citenamefont
  {Ilani}}]{RN20}%
  \BibitemOpen
  \bibfield  {author} {\bibinfo {author} {\bibfnamefont {D.~R.}\ \bibnamefont
  {Klein}}, \bibinfo {author} {\bibfnamefont {U.}~\bibnamefont {Zondiner}},
  \bibinfo {author} {\bibfnamefont {A.}~\bibnamefont {Keren}}, \bibinfo
  {author} {\bibfnamefont {J.}~\bibnamefont {Birkbeck}}, \bibinfo {author}
  {\bibfnamefont {A.}~\bibnamefont {Inbar}}, \bibinfo {author} {\bibfnamefont
  {J.}~\bibnamefont {Xiao}}, \bibinfo {author} {\bibfnamefont {M.}~\bibnamefont
  {Sidorova}}, \bibinfo {author} {\bibfnamefont {M.~M.~A.}\ \bibnamefont
  {Ezzi}}, \bibinfo {author} {\bibfnamefont {L.}~\bibnamefont {Peng}}, \bibinfo
  {author} {\bibfnamefont {K.}~\bibnamefont {Watanabe}}, \bibinfo {author}
  {\bibfnamefont {T.}~\bibnamefont {Taniguchi}}, \bibinfo {author}
  {\bibfnamefont {S.}~\bibnamefont {Adam}},\ and\ \bibinfo {author}
  {\bibfnamefont {S.}~\bibnamefont {Ilani}},\ }\href
  {https://arxiv.org/abs/2410.22277} {\bibinfo {title} {Imaging the sub-moir\'e
  potential landscape using an atomic single electron transistor}} (\bibinfo
  {year} {2024}),\ \Eprint {https://arxiv.org/abs/2410.22277} {arXiv:2410.22277
  [cond-mat.mes-hall]} \BibitemShut {NoStop}%
\bibitem [{\citenamefont {Morello}\ \emph {et~al.}(2010)\citenamefont
  {Morello}, \citenamefont {Pla}, \citenamefont {Zwanenburg}, \citenamefont
  {Chan}, \citenamefont {Tan}, \citenamefont {Huebl}, \citenamefont
  {Möttönen}, \citenamefont {Nugroho}, \citenamefont {Yang}, \citenamefont
  {van Donkelaar}, \citenamefont {Alves}, \citenamefont {Jamieson},
  \citenamefont {Escott}, \citenamefont {Hollenberg}, \citenamefont {Clark},\
  and\ \citenamefont {Dzurak}}]{RN21}%
  \BibitemOpen
  \bibfield  {author} {\bibinfo {author} {\bibfnamefont {A.}~\bibnamefont
  {Morello}}, \bibinfo {author} {\bibfnamefont {J.~J.}\ \bibnamefont {Pla}},
  \bibinfo {author} {\bibfnamefont {F.~A.}\ \bibnamefont {Zwanenburg}},
  \bibinfo {author} {\bibfnamefont {K.~W.}\ \bibnamefont {Chan}}, \bibinfo
  {author} {\bibfnamefont {K.~Y.}\ \bibnamefont {Tan}}, \bibinfo {author}
  {\bibfnamefont {H.}~\bibnamefont {Huebl}}, \bibinfo {author} {\bibfnamefont
  {M.}~\bibnamefont {Möttönen}}, \bibinfo {author} {\bibfnamefont {C.~D.}\
  \bibnamefont {Nugroho}}, \bibinfo {author} {\bibfnamefont {C.}~\bibnamefont
  {Yang}}, \bibinfo {author} {\bibfnamefont {J.~A.}\ \bibnamefont {van
  Donkelaar}}, \bibinfo {author} {\bibfnamefont {A.~D.~C.}\ \bibnamefont
  {Alves}}, \bibinfo {author} {\bibfnamefont {D.~N.}\ \bibnamefont {Jamieson}},
  \bibinfo {author} {\bibfnamefont {C.~C.}\ \bibnamefont {Escott}}, \bibinfo
  {author} {\bibfnamefont {L.~C.~L.}\ \bibnamefont {Hollenberg}}, \bibinfo
  {author} {\bibfnamefont {R.~G.}\ \bibnamefont {Clark}},\ and\ \bibinfo
  {author} {\bibfnamefont {A.~S.}\ \bibnamefont {Dzurak}},\ }\bibfield  {title}
  {\bibinfo {title} {Single-shot readout of an electron spin in silicon},\
  }\href {https://doi.org/10.1038/nature09392} {\bibfield  {journal} {\bibinfo
  {journal} {Nature}\ }\textbf {\bibinfo {volume} {467}},\ \bibinfo {pages}
  {687} (\bibinfo {year} {2010})}\BibitemShut {NoStop}%
\bibitem [{\citenamefont {Oakes}\ \emph {et~al.}(2023)\citenamefont {Oakes},
  \citenamefont {Ciriano-Tejel}, \citenamefont {Wise}, \citenamefont {Fogarty},
  \citenamefont {Lundberg}, \citenamefont {Lain\'e}, \citenamefont {Schaal},
  \citenamefont {Martins}, \citenamefont {Ibberson}, \citenamefont {Hutin},
  \citenamefont {Bertrand}, \citenamefont {Stelmashenko}, \citenamefont
  {Robinson}, \citenamefont {Ibberson}, \citenamefont {Hashim}, \citenamefont
  {Siddiqi}, \citenamefont {Lee}, \citenamefont {Vinet}, \citenamefont {Smith},
  \citenamefont {Morton},\ and\ \citenamefont {Gonzalez-Zalba}}]{RN22}%
  \BibitemOpen
  \bibfield  {author} {\bibinfo {author} {\bibfnamefont {G.~A.}\ \bibnamefont
  {Oakes}}, \bibinfo {author} {\bibfnamefont {V.~N.}\ \bibnamefont
  {Ciriano-Tejel}}, \bibinfo {author} {\bibfnamefont {D.~F.}\ \bibnamefont
  {Wise}}, \bibinfo {author} {\bibfnamefont {M.~A.}\ \bibnamefont {Fogarty}},
  \bibinfo {author} {\bibfnamefont {T.}~\bibnamefont {Lundberg}}, \bibinfo
  {author} {\bibfnamefont {C.}~\bibnamefont {Lain\'e}}, \bibinfo {author}
  {\bibfnamefont {S.}~\bibnamefont {Schaal}}, \bibinfo {author} {\bibfnamefont
  {F.}~\bibnamefont {Martins}}, \bibinfo {author} {\bibfnamefont {D.~J.}\
  \bibnamefont {Ibberson}}, \bibinfo {author} {\bibfnamefont {L.}~\bibnamefont
  {Hutin}}, \bibinfo {author} {\bibfnamefont {B.}~\bibnamefont {Bertrand}},
  \bibinfo {author} {\bibfnamefont {N.}~\bibnamefont {Stelmashenko}}, \bibinfo
  {author} {\bibfnamefont {J.~W.~A.}\ \bibnamefont {Robinson}}, \bibinfo
  {author} {\bibfnamefont {L.}~\bibnamefont {Ibberson}}, \bibinfo {author}
  {\bibfnamefont {A.}~\bibnamefont {Hashim}}, \bibinfo {author} {\bibfnamefont
  {I.}~\bibnamefont {Siddiqi}}, \bibinfo {author} {\bibfnamefont
  {A.}~\bibnamefont {Lee}}, \bibinfo {author} {\bibfnamefont {M.}~\bibnamefont
  {Vinet}}, \bibinfo {author} {\bibfnamefont {C.~G.}\ \bibnamefont {Smith}},
  \bibinfo {author} {\bibfnamefont {J.~J.~L.}\ \bibnamefont {Morton}},\ and\
  \bibinfo {author} {\bibfnamefont {M.~F.}\ \bibnamefont {Gonzalez-Zalba}},\
  }\bibfield  {title} {\bibinfo {title} {Fast high-fidelity single-shot readout
  of spins in silicon using a single-electron box},\ }\href
  {https://doi.org/10.1103/PhysRevX.13.011023} {\bibfield  {journal} {\bibinfo
  {journal} {Physical Review X}\ }\textbf {\bibinfo {volume} {13}},\ \bibinfo
  {pages} {011023} (\bibinfo {year} {2023})}\BibitemShut {NoStop}%
\bibitem [{\citenamefont {Hanson}\ \emph {et~al.}(2005)\citenamefont {Hanson},
  \citenamefont {van Beveren}, \citenamefont {Vink}, \citenamefont {Elzerman},
  \citenamefont {Naber}, \citenamefont {Koppens}, \citenamefont {Kouwenhoven},\
  and\ \citenamefont {Vandersypen}}]{RN23}%
  \BibitemOpen
  \bibfield  {author} {\bibinfo {author} {\bibfnamefont {R.}~\bibnamefont
  {Hanson}}, \bibinfo {author} {\bibfnamefont {L.~H.~W.}\ \bibnamefont {van
  Beveren}}, \bibinfo {author} {\bibfnamefont {I.~T.}\ \bibnamefont {Vink}},
  \bibinfo {author} {\bibfnamefont {J.~M.}\ \bibnamefont {Elzerman}}, \bibinfo
  {author} {\bibfnamefont {W.~J.~M.}\ \bibnamefont {Naber}}, \bibinfo {author}
  {\bibfnamefont {F.~H.~L.}\ \bibnamefont {Koppens}}, \bibinfo {author}
  {\bibfnamefont {L.~P.}\ \bibnamefont {Kouwenhoven}},\ and\ \bibinfo {author}
  {\bibfnamefont {L.~M.~K.}\ \bibnamefont {Vandersypen}},\ }\bibfield  {title}
  {\bibinfo {title} {Single-shot readout of electron spin states in a quantum
  dot using spin-dependent tunnel rates},\ }\href
  {https://doi.org/10.1103/PhysRevLett.94.196802} {\bibfield  {journal}
  {\bibinfo  {journal} {Physical Review Letters}\ }\textbf {\bibinfo {volume}
  {94}},\ \bibinfo {pages} {196802} (\bibinfo {year} {2005})}\BibitemShut
  {NoStop}%
\bibitem [{\citenamefont {Godfrin}\ \emph {et~al.}(2017)\citenamefont
  {Godfrin}, \citenamefont {Thiele}, \citenamefont {Ferhat}, \citenamefont
  {Klyatskaya}, \citenamefont {Ruben}, \citenamefont {Wernsdorfer},\ and\
  \citenamefont {Balestro}}]{RN24}%
  \BibitemOpen
  \bibfield  {author} {\bibinfo {author} {\bibfnamefont {C.}~\bibnamefont
  {Godfrin}}, \bibinfo {author} {\bibfnamefont {S.}~\bibnamefont {Thiele}},
  \bibinfo {author} {\bibfnamefont {A.}~\bibnamefont {Ferhat}}, \bibinfo
  {author} {\bibfnamefont {S.}~\bibnamefont {Klyatskaya}}, \bibinfo {author}
  {\bibfnamefont {M.}~\bibnamefont {Ruben}}, \bibinfo {author} {\bibfnamefont
  {W.}~\bibnamefont {Wernsdorfer}},\ and\ \bibinfo {author} {\bibfnamefont
  {F.}~\bibnamefont {Balestro}},\ }\bibfield  {title} {\bibinfo {title}
  {Electrical read-out of a single spin using an exchange-coupled quantum
  dot},\ }\href {https://doi.org/10.1021/acsnano.7b00451} {\bibfield  {journal}
  {\bibinfo  {journal} {ACS Nano}\ }\textbf {\bibinfo {volume} {11}},\ \bibinfo
  {pages} {3984} (\bibinfo {year} {2017})}\BibitemShut {NoStop}%
\bibitem [{\citenamefont {Gulka}\ \emph {et~al.}(2021)\citenamefont {Gulka},
  \citenamefont {Wirtitsch}, \citenamefont {Ivády}, \citenamefont {Vodnik},
  \citenamefont {Hruby}, \citenamefont {Magchiels}, \citenamefont {Bourgeois},
  \citenamefont {Gali}, \citenamefont {Trupke},\ and\ \citenamefont
  {Nesladek}}]{RN25}%
  \BibitemOpen
  \bibfield  {author} {\bibinfo {author} {\bibfnamefont {M.}~\bibnamefont
  {Gulka}}, \bibinfo {author} {\bibfnamefont {D.}~\bibnamefont {Wirtitsch}},
  \bibinfo {author} {\bibfnamefont {V.}~\bibnamefont {Ivády}}, \bibinfo
  {author} {\bibfnamefont {J.}~\bibnamefont {Vodnik}}, \bibinfo {author}
  {\bibfnamefont {J.}~\bibnamefont {Hruby}}, \bibinfo {author} {\bibfnamefont
  {G.}~\bibnamefont {Magchiels}}, \bibinfo {author} {\bibfnamefont
  {E.}~\bibnamefont {Bourgeois}}, \bibinfo {author} {\bibfnamefont
  {A.}~\bibnamefont {Gali}}, \bibinfo {author} {\bibfnamefont {M.}~\bibnamefont
  {Trupke}},\ and\ \bibinfo {author} {\bibfnamefont {M.}~\bibnamefont
  {Nesladek}},\ }\bibfield  {title} {\bibinfo {title} {Room-temperature control
  and electrical readout of individual nitrogen-vacancy nuclear spins},\ }\href
  {https://doi.org/10.1038/s41467-021-24494-x} {\bibfield  {journal} {\bibinfo
  {journal} {Nature Communications}\ }\textbf {\bibinfo {volume} {12}},\
  \bibinfo {pages} {4421} (\bibinfo {year} {2021})}\BibitemShut {NoStop}%
\bibitem [{\citenamefont {Siyushev}\ \emph {et~al.}(2019)\citenamefont
  {Siyushev}, \citenamefont {Nesladek}, \citenamefont {Bourgeois},
  \citenamefont {Gulka}, \citenamefont {Hruby}, \citenamefont {Yamamoto},
  \citenamefont {Trupke}, \citenamefont {Teraji}, \citenamefont {Isoya},\ and\
  \citenamefont {Jelezko}}]{RN26}%
  \BibitemOpen
  \bibfield  {author} {\bibinfo {author} {\bibfnamefont {P.}~\bibnamefont
  {Siyushev}}, \bibinfo {author} {\bibfnamefont {M.}~\bibnamefont {Nesladek}},
  \bibinfo {author} {\bibfnamefont {E.}~\bibnamefont {Bourgeois}}, \bibinfo
  {author} {\bibfnamefont {M.}~\bibnamefont {Gulka}}, \bibinfo {author}
  {\bibfnamefont {J.}~\bibnamefont {Hruby}}, \bibinfo {author} {\bibfnamefont
  {T.}~\bibnamefont {Yamamoto}}, \bibinfo {author} {\bibfnamefont
  {M.}~\bibnamefont {Trupke}}, \bibinfo {author} {\bibfnamefont
  {T.}~\bibnamefont {Teraji}}, \bibinfo {author} {\bibfnamefont
  {J.}~\bibnamefont {Isoya}},\ and\ \bibinfo {author} {\bibfnamefont
  {F.}~\bibnamefont {Jelezko}},\ }\bibfield  {title} {\bibinfo {title}
  {Photoelectrical imaging and coherent spin-state readout of single
  nitrogen-vacancy centers in diamond},\ }\href
  {https://doi.org/10.1126/science.aav2789} {\bibfield  {journal} {\bibinfo
  {journal} {Science}\ }\textbf {\bibinfo {volume} {363}},\ \bibinfo {pages}
  {728} (\bibinfo {year} {2019})}\BibitemShut {NoStop}%
\bibitem [{\citenamefont {Bourgeois}\ \emph {et~al.}(2015)\citenamefont
  {Bourgeois}, \citenamefont {Jarmola}, \citenamefont {Siyushev}, \citenamefont
  {Gulka}, \citenamefont {Hruby}, \citenamefont {Jelezko}, \citenamefont
  {Budker},\ and\ \citenamefont {Nesladek}}]{RN27}%
  \BibitemOpen
  \bibfield  {author} {\bibinfo {author} {\bibfnamefont {E.}~\bibnamefont
  {Bourgeois}}, \bibinfo {author} {\bibfnamefont {A.}~\bibnamefont {Jarmola}},
  \bibinfo {author} {\bibfnamefont {P.}~\bibnamefont {Siyushev}}, \bibinfo
  {author} {\bibfnamefont {M.}~\bibnamefont {Gulka}}, \bibinfo {author}
  {\bibfnamefont {J.}~\bibnamefont {Hruby}}, \bibinfo {author} {\bibfnamefont
  {F.}~\bibnamefont {Jelezko}}, \bibinfo {author} {\bibfnamefont
  {D.}~\bibnamefont {Budker}},\ and\ \bibinfo {author} {\bibfnamefont
  {M.}~\bibnamefont {Nesladek}},\ }\bibfield  {title} {\bibinfo {title}
  {Photoelectric detection of electron spin resonance of nitrogen-vacancy
  centres in diamond},\ }\href {https://doi.org/10.1038/ncomms9577} {\bibfield
  {journal} {\bibinfo  {journal} {Nature Communications}\ }\textbf {\bibinfo
  {volume} {6}},\ \bibinfo {pages} {8577} (\bibinfo {year} {2015})}\BibitemShut
  {NoStop}%
\bibitem [{\citenamefont {Hrubesch}\ \emph {et~al.}(2017)\citenamefont
  {Hrubesch}, \citenamefont {Braunbeck}, \citenamefont {Stutzmann},
  \citenamefont {Reinhard},\ and\ \citenamefont {Brandt}}]{RN28}%
  \BibitemOpen
  \bibfield  {author} {\bibinfo {author} {\bibfnamefont {F.~M.}\ \bibnamefont
  {Hrubesch}}, \bibinfo {author} {\bibfnamefont {G.}~\bibnamefont {Braunbeck}},
  \bibinfo {author} {\bibfnamefont {M.}~\bibnamefont {Stutzmann}}, \bibinfo
  {author} {\bibfnamefont {F.}~\bibnamefont {Reinhard}},\ and\ \bibinfo
  {author} {\bibfnamefont {M.~S.}\ \bibnamefont {Brandt}},\ }\bibfield  {title}
  {\bibinfo {title} {Efficient electrical spin readout of
  ${\mathrm{nv}}^{\ensuremath{-}}$ centers in diamond},\ }\href
  {https://doi.org/10.1103/PhysRevLett.118.037601} {\bibfield  {journal}
  {\bibinfo  {journal} {Physical Review Letters}\ }\textbf {\bibinfo {volume}
  {118}},\ \bibinfo {pages} {037601} (\bibinfo {year} {2017})}\BibitemShut
  {NoStop}%
\bibitem [{\citenamefont {Niethammer}\ \emph {et~al.}(2019)\citenamefont
  {Niethammer}, \citenamefont {Widmann}, \citenamefont {Rendler}, \citenamefont
  {Morioka}, \citenamefont {Chen}, \citenamefont {Stöhr}, \citenamefont
  {Hassan}, \citenamefont {Onoda}, \citenamefont {Ohshima}, \citenamefont
  {Lee}, \citenamefont {Mukherjee}, \citenamefont {Isoya}, \citenamefont
  {Son},\ and\ \citenamefont {Wrachtrup}}]{RN29}%
  \BibitemOpen
  \bibfield  {author} {\bibinfo {author} {\bibfnamefont {M.}~\bibnamefont
  {Niethammer}}, \bibinfo {author} {\bibfnamefont {M.}~\bibnamefont {Widmann}},
  \bibinfo {author} {\bibfnamefont {T.}~\bibnamefont {Rendler}}, \bibinfo
  {author} {\bibfnamefont {N.}~\bibnamefont {Morioka}}, \bibinfo {author}
  {\bibfnamefont {Y.-C.}\ \bibnamefont {Chen}}, \bibinfo {author}
  {\bibfnamefont {R.}~\bibnamefont {Stöhr}}, \bibinfo {author} {\bibfnamefont
  {J.~U.}\ \bibnamefont {Hassan}}, \bibinfo {author} {\bibfnamefont
  {S.}~\bibnamefont {Onoda}}, \bibinfo {author} {\bibfnamefont
  {T.}~\bibnamefont {Ohshima}}, \bibinfo {author} {\bibfnamefont {S.-Y.}\
  \bibnamefont {Lee}}, \bibinfo {author} {\bibfnamefont {A.}~\bibnamefont
  {Mukherjee}}, \bibinfo {author} {\bibfnamefont {J.}~\bibnamefont {Isoya}},
  \bibinfo {author} {\bibfnamefont {N.~T.}\ \bibnamefont {Son}},\ and\ \bibinfo
  {author} {\bibfnamefont {J.}~\bibnamefont {Wrachtrup}},\ }\bibfield  {title}
  {\bibinfo {title} {Coherent electrical readout of defect spins in silicon
  carbide by photo-ionization at ambient conditions},\ }\href
  {https://doi.org/10.1038/s41467-019-13545-z} {\bibfield  {journal} {\bibinfo
  {journal} {Nature Communications}\ }\textbf {\bibinfo {volume} {10}},\
  \bibinfo {pages} {5569} (\bibinfo {year} {2019})}\BibitemShut {NoStop}%
\bibitem [{\citenamefont {Lew}\ \emph {et~al.}(2024)\citenamefont {Lew},
  \citenamefont {Sewani}, \citenamefont {Iwamoto}, \citenamefont {Ohshima},
  \citenamefont {McCallum},\ and\ \citenamefont {Johnson}}]{RN30}%
  \BibitemOpen
  \bibfield  {author} {\bibinfo {author} {\bibfnamefont {C.~T.-K.}\
  \bibnamefont {Lew}}, \bibinfo {author} {\bibfnamefont {V.~K.}\ \bibnamefont
  {Sewani}}, \bibinfo {author} {\bibfnamefont {N.}~\bibnamefont {Iwamoto}},
  \bibinfo {author} {\bibfnamefont {T.}~\bibnamefont {Ohshima}}, \bibinfo
  {author} {\bibfnamefont {J.~C.}\ \bibnamefont {McCallum}},\ and\ \bibinfo
  {author} {\bibfnamefont {B.~C.}\ \bibnamefont {Johnson}},\ }\bibfield
  {title} {\bibinfo {title} {All-electrical readout of coherently controlled
  spins in silicon carbide},\ }\href
  {https://doi.org/10.1103/PhysRevLett.132.146902} {\bibfield  {journal}
  {\bibinfo  {journal} {Physical Review Letters}\ }\textbf {\bibinfo {volume}
  {132}},\ \bibinfo {pages} {146902} (\bibinfo {year} {2024})}\BibitemShut
  {NoStop}%
\bibitem [{\citenamefont {Mu}\ \emph {et~al.}(2022)\citenamefont {Mu},
  \citenamefont {Cai}, \citenamefont {Chen}, \citenamefont {Kenny},
  \citenamefont {Jiang}, \citenamefont {Ru}, \citenamefont {Lyu}, \citenamefont
  {Koh}, \citenamefont {Liu}, \citenamefont {Aharonovich},\ and\ \citenamefont
  {Gao}}]{RN31}%
  \BibitemOpen
  \bibfield  {author} {\bibinfo {author} {\bibfnamefont {Z.}~\bibnamefont
  {Mu}}, \bibinfo {author} {\bibfnamefont {H.}~\bibnamefont {Cai}}, \bibinfo
  {author} {\bibfnamefont {D.}~\bibnamefont {Chen}}, \bibinfo {author}
  {\bibfnamefont {J.}~\bibnamefont {Kenny}}, \bibinfo {author} {\bibfnamefont
  {Z.}~\bibnamefont {Jiang}}, \bibinfo {author} {\bibfnamefont
  {S.}~\bibnamefont {Ru}}, \bibinfo {author} {\bibfnamefont {X.}~\bibnamefont
  {Lyu}}, \bibinfo {author} {\bibfnamefont {T.~S.}\ \bibnamefont {Koh}},
  \bibinfo {author} {\bibfnamefont {X.}~\bibnamefont {Liu}}, \bibinfo {author}
  {\bibfnamefont {I.}~\bibnamefont {Aharonovich}},\ and\ \bibinfo {author}
  {\bibfnamefont {W.}~\bibnamefont {Gao}},\ }\bibfield  {title} {\bibinfo
  {title} {Excited-state optically detected magnetic resonance of spin defects
  in hexagonal boron nitride},\ }\href
  {https://doi.org/10.1103/PhysRevLett.128.216402} {\bibfield  {journal}
  {\bibinfo  {journal} {Physical Review Letters}\ }\textbf {\bibinfo {volume}
  {128}},\ \bibinfo {pages} {216402} (\bibinfo {year} {2022})}\BibitemShut
  {NoStop}%
\bibitem [{\citenamefont {Mathur}\ \emph {et~al.}(2022)\citenamefont {Mathur},
  \citenamefont {Mukherjee}, \citenamefont {Gao}, \citenamefont {Luo},
  \citenamefont {McCullian}, \citenamefont {Li}, \citenamefont {Vamivakas},\
  and\ \citenamefont {Fuchs}}]{RN32}%
  \BibitemOpen
  \bibfield  {author} {\bibinfo {author} {\bibfnamefont {N.}~\bibnamefont
  {Mathur}}, \bibinfo {author} {\bibfnamefont {A.}~\bibnamefont {Mukherjee}},
  \bibinfo {author} {\bibfnamefont {X.}~\bibnamefont {Gao}}, \bibinfo {author}
  {\bibfnamefont {J.}~\bibnamefont {Luo}}, \bibinfo {author} {\bibfnamefont
  {B.~A.}\ \bibnamefont {McCullian}}, \bibinfo {author} {\bibfnamefont
  {T.}~\bibnamefont {Li}}, \bibinfo {author} {\bibfnamefont {A.~N.}\
  \bibnamefont {Vamivakas}},\ and\ \bibinfo {author} {\bibfnamefont {G.~D.}\
  \bibnamefont {Fuchs}},\ }\bibfield  {title} {\bibinfo {title} {Excited-state
  spin-resonance spectroscopy of $\mathrm{V_B^-}$ defect centers in hexagonal
  boron nitride},\ }\href {https://doi.org/10.1038/s41467-022-30772-z}
  {\bibfield  {journal} {\bibinfo  {journal} {Nature Communications}\ }\textbf
  {\bibinfo {volume} {13}},\ \bibinfo {pages} {3233} (\bibinfo {year}
  {2022})}\BibitemShut {NoStop}%
\bibitem [{\citenamefont {Yu}\ \emph {et~al.}(2022)\citenamefont {Yu},
  \citenamefont {Sun}, \citenamefont {Wang}, \citenamefont {Zhang},
  \citenamefont {Ye}, \citenamefont {Zhou}, \citenamefont {Liu}, \citenamefont
  {Wang}, \citenamefont {Shi}, \citenamefont {Wang},\ and\ \citenamefont
  {Du}}]{RN33}%
  \BibitemOpen
  \bibfield  {author} {\bibinfo {author} {\bibfnamefont {P.}~\bibnamefont
  {Yu}}, \bibinfo {author} {\bibfnamefont {H.}~\bibnamefont {Sun}}, \bibinfo
  {author} {\bibfnamefont {M.}~\bibnamefont {Wang}}, \bibinfo {author}
  {\bibfnamefont {T.}~\bibnamefont {Zhang}}, \bibinfo {author} {\bibfnamefont
  {X.}~\bibnamefont {Ye}}, \bibinfo {author} {\bibfnamefont {J.}~\bibnamefont
  {Zhou}}, \bibinfo {author} {\bibfnamefont {H.}~\bibnamefont {Liu}}, \bibinfo
  {author} {\bibfnamefont {C.-J.}\ \bibnamefont {Wang}}, \bibinfo {author}
  {\bibfnamefont {F.}~\bibnamefont {Shi}}, \bibinfo {author} {\bibfnamefont
  {Y.}~\bibnamefont {Wang}},\ and\ \bibinfo {author} {\bibfnamefont
  {J.}~\bibnamefont {Du}},\ }\bibfield  {title} {\bibinfo {title}
  {Excited-state spectroscopy of spin defects in hexagonal boron nitride},\
  }\href {https://doi.org/10.1021/acs.nanolett.1c04841} {\bibfield  {journal}
  {\bibinfo  {journal} {Nano Letters}\ }\textbf {\bibinfo {volume} {22}},\
  \bibinfo {pages} {3545} (\bibinfo {year} {2022})}\BibitemShut {NoStop}%
\bibitem [{Sup()}]{Supp_info}%
  \BibitemOpen
  \bibfield  {title} {\bibinfo {title} {See supplemental material at
  https://——/—–, which includes refs. [26, 31, 38], for additional
  information about the experimental methods and a detailed discussion of
  experimental and simulation results}}\href@noop {} {\ }\BibitemShut
  {NoStop}%
\bibitem [{\citenamefont {Rizzato}\ \emph {et~al.}(2023)\citenamefont
  {Rizzato}, \citenamefont {Schalk}, \citenamefont {Mohr}, \citenamefont
  {Hermann}, \citenamefont {Leibold}, \citenamefont {Bruckmaier}, \citenamefont
  {Salvitti}, \citenamefont {Qian}, \citenamefont {Ji}, \citenamefont
  {Astakhov} \emph {et~al.}}]{RN34}%
  \BibitemOpen
  \bibfield  {author} {\bibinfo {author} {\bibfnamefont {R.}~\bibnamefont
  {Rizzato}}, \bibinfo {author} {\bibfnamefont {M.}~\bibnamefont {Schalk}},
  \bibinfo {author} {\bibfnamefont {S.}~\bibnamefont {Mohr}}, \bibinfo {author}
  {\bibfnamefont {J.~C.}\ \bibnamefont {Hermann}}, \bibinfo {author}
  {\bibfnamefont {J.~P.}\ \bibnamefont {Leibold}}, \bibinfo {author}
  {\bibfnamefont {F.}~\bibnamefont {Bruckmaier}}, \bibinfo {author}
  {\bibfnamefont {G.}~\bibnamefont {Salvitti}}, \bibinfo {author}
  {\bibfnamefont {C.}~\bibnamefont {Qian}}, \bibinfo {author} {\bibfnamefont
  {P.}~\bibnamefont {Ji}}, \bibinfo {author} {\bibfnamefont {G.~V.}\
  \bibnamefont {Astakhov}}, \emph {et~al.},\ }\bibfield  {title} {\bibinfo
  {title} {Extending the coherence of spin defects in hbn enables advanced
  qubit control and quantum sensing},\ }\href
  {https://doi.org/https://doi.org/10.1038/s41467-023-40473-w} {\bibfield
  {journal} {\bibinfo  {journal} {Nature Communications}\ }\textbf {\bibinfo
  {volume} {14}},\ \bibinfo {pages} {5089} (\bibinfo {year}
  {2023})}\BibitemShut {NoStop}%
\bibitem [{\citenamefont {Gong}\ \emph {et~al.}(2024)\citenamefont {Gong},
  \citenamefont {Du}, \citenamefont {Janzen}, \citenamefont {Liu},
  \citenamefont {Liu}, \citenamefont {He}, \citenamefont {Ye}, \citenamefont
  {Li}, \citenamefont {Yao}, \citenamefont {Edgar}, \citenamefont {Henriksen},\
  and\ \citenamefont {Zu}}]{RN35}%
  \BibitemOpen
  \bibfield  {author} {\bibinfo {author} {\bibfnamefont {R.}~\bibnamefont
  {Gong}}, \bibinfo {author} {\bibfnamefont {X.}~\bibnamefont {Du}}, \bibinfo
  {author} {\bibfnamefont {E.}~\bibnamefont {Janzen}}, \bibinfo {author}
  {\bibfnamefont {V.}~\bibnamefont {Liu}}, \bibinfo {author} {\bibfnamefont
  {Z.}~\bibnamefont {Liu}}, \bibinfo {author} {\bibfnamefont {G.}~\bibnamefont
  {He}}, \bibinfo {author} {\bibfnamefont {B.}~\bibnamefont {Ye}}, \bibinfo
  {author} {\bibfnamefont {T.}~\bibnamefont {Li}}, \bibinfo {author}
  {\bibfnamefont {N.~Y.}\ \bibnamefont {Yao}}, \bibinfo {author} {\bibfnamefont
  {J.~H.}\ \bibnamefont {Edgar}}, \bibinfo {author} {\bibfnamefont {E.~A.}\
  \bibnamefont {Henriksen}},\ and\ \bibinfo {author} {\bibfnamefont
  {C.}~\bibnamefont {Zu}},\ }\bibfield  {title} {\bibinfo {title} {Isotope
  engineering for spin defects in van der waals materials},\ }\href
  {https://doi.org/10.1038/s41467-023-44494-3} {\bibfield  {journal} {\bibinfo
  {journal} {Nature Communications}\ }\textbf {\bibinfo {volume} {15}},\
  \bibinfo {pages} {104} (\bibinfo {year} {2024})}\BibitemShut {NoStop}%
\bibitem [{\citenamefont {Shim}\ \emph {et~al.}(2012)\citenamefont {Shim},
  \citenamefont {Niemeyer}, \citenamefont {Zhang},\ and\ \citenamefont
  {Suter}}]{RN36}%
  \BibitemOpen
  \bibfield  {author} {\bibinfo {author} {\bibfnamefont {J.}~\bibnamefont
  {Shim}}, \bibinfo {author} {\bibfnamefont {I.}~\bibnamefont {Niemeyer}},
  \bibinfo {author} {\bibfnamefont {J.}~\bibnamefont {Zhang}},\ and\ \bibinfo
  {author} {\bibfnamefont {D.}~\bibnamefont {Suter}},\ }\bibfield  {title}
  {\bibinfo {title} {Robust dynamical decoupling for arbitrary quantum states
  of a single nv center in diamond},\ }\href
  {https://doi.org/10.1209/0295-5075/99/40004} {\bibfield  {journal} {\bibinfo
  {journal} {Europhysics Letters}\ }\textbf {\bibinfo {volume} {99}},\ \bibinfo
  {pages} {40004} (\bibinfo {year} {2012})}\BibitemShut {NoStop}%
\bibitem [{\citenamefont {Ru}\ \emph {et~al.}(2024)\citenamefont {Ru},
  \citenamefont {Jiang}, \citenamefont {Liang}, \citenamefont {Kenny},
  \citenamefont {Cai}, \citenamefont {Lyu}, \citenamefont {Cernansky},
  \citenamefont {Zhou}, \citenamefont {Yang}, \citenamefont {Watanabe},
  \citenamefont {Taniguch}, \citenamefont {Li}, \citenamefont {Koh},
  \citenamefont {Liu}, \citenamefont {Jelezko}, \citenamefont {Bettiol},\ and\
  \citenamefont {Gao}}]{RN37}%
  \BibitemOpen
  \bibfield  {author} {\bibinfo {author} {\bibfnamefont {S.}~\bibnamefont
  {Ru}}, \bibinfo {author} {\bibfnamefont {Z.}~\bibnamefont {Jiang}}, \bibinfo
  {author} {\bibfnamefont {H.}~\bibnamefont {Liang}}, \bibinfo {author}
  {\bibfnamefont {J.}~\bibnamefont {Kenny}}, \bibinfo {author} {\bibfnamefont
  {H.}~\bibnamefont {Cai}}, \bibinfo {author} {\bibfnamefont {X.}~\bibnamefont
  {Lyu}}, \bibinfo {author} {\bibfnamefont {R.}~\bibnamefont {Cernansky}},
  \bibinfo {author} {\bibfnamefont {F.}~\bibnamefont {Zhou}}, \bibinfo {author}
  {\bibfnamefont {Y.}~\bibnamefont {Yang}}, \bibinfo {author} {\bibfnamefont
  {K.}~\bibnamefont {Watanabe}}, \bibinfo {author} {\bibfnamefont
  {T.}~\bibnamefont {Taniguch}}, \bibinfo {author} {\bibfnamefont
  {F.}~\bibnamefont {Li}}, \bibinfo {author} {\bibfnamefont {T.~S.}\
  \bibnamefont {Koh}}, \bibinfo {author} {\bibfnamefont {X.}~\bibnamefont
  {Liu}}, \bibinfo {author} {\bibfnamefont {F.}~\bibnamefont {Jelezko}},
  \bibinfo {author} {\bibfnamefont {A.~A.}\ \bibnamefont {Bettiol}},\ and\
  \bibinfo {author} {\bibfnamefont {W.}~\bibnamefont {Gao}},\ }\bibfield
  {title} {\bibinfo {title} {Robust nuclear spin polarization via ground-state
  level anticrossing of boron vacancy defects in hexagonal boron nitride},\
  }\href {https://doi.org/10.1103/PhysRevLett.132.266801} {\bibfield  {journal}
  {\bibinfo  {journal} {Physical Review Letters}\ }\textbf {\bibinfo {volume}
  {132}},\ \bibinfo {pages} {266801} (\bibinfo {year} {2024})}\BibitemShut
  {NoStop}%
\bibitem [{\citenamefont {Gao}\ \emph {et~al.}(2022)\citenamefont {Gao},
  \citenamefont {Vaidya}, \citenamefont {Li}, \citenamefont {Ju}, \citenamefont
  {Jiang}, \citenamefont {Xu}, \citenamefont {Allcca}, \citenamefont {Shen},
  \citenamefont {Taniguchi}, \citenamefont {Watanabe}, \citenamefont {Bhave},
  \citenamefont {Chen}, \citenamefont {Ping},\ and\ \citenamefont {Li}}]{RN38}%
  \BibitemOpen
  \bibfield  {author} {\bibinfo {author} {\bibfnamefont {X.}~\bibnamefont
  {Gao}}, \bibinfo {author} {\bibfnamefont {S.}~\bibnamefont {Vaidya}},
  \bibinfo {author} {\bibfnamefont {K.}~\bibnamefont {Li}}, \bibinfo {author}
  {\bibfnamefont {P.}~\bibnamefont {Ju}}, \bibinfo {author} {\bibfnamefont
  {B.}~\bibnamefont {Jiang}}, \bibinfo {author} {\bibfnamefont
  {Z.}~\bibnamefont {Xu}}, \bibinfo {author} {\bibfnamefont {A.~E.~L.}\
  \bibnamefont {Allcca}}, \bibinfo {author} {\bibfnamefont {K.}~\bibnamefont
  {Shen}}, \bibinfo {author} {\bibfnamefont {T.}~\bibnamefont {Taniguchi}},
  \bibinfo {author} {\bibfnamefont {K.}~\bibnamefont {Watanabe}}, \bibinfo
  {author} {\bibfnamefont {S.~A.}\ \bibnamefont {Bhave}}, \bibinfo {author}
  {\bibfnamefont {Y.~P.}\ \bibnamefont {Chen}}, \bibinfo {author}
  {\bibfnamefont {Y.}~\bibnamefont {Ping}},\ and\ \bibinfo {author}
  {\bibfnamefont {T.}~\bibnamefont {Li}},\ }\bibfield  {title} {\bibinfo
  {title} {Nuclear spin polarization and control in hexagonal boron nitride},\
  }\href {https://doi.org/10.1038/s41563-022-01329-8} {\bibfield  {journal}
  {\bibinfo  {journal} {Nature Materials}\ }\textbf {\bibinfo {volume} {21}},\
  \bibinfo {pages} {1024} (\bibinfo {year} {2022})}\BibitemShut {NoStop}%
\bibitem [{\citenamefont {Clua-Provost}\ \emph {et~al.}(2023)\citenamefont
  {Clua-Provost}, \citenamefont {Durand}, \citenamefont {Mu}, \citenamefont
  {Rastoin}, \citenamefont {Frauni\'e}, \citenamefont {Janzen}, \citenamefont
  {Schutte}, \citenamefont {Edgar}, \citenamefont {Seine}, \citenamefont
  {Claverie}, \citenamefont {Marie}, \citenamefont {Robert}, \citenamefont
  {Gil}, \citenamefont {Cassabois},\ and\ \citenamefont {Jacques}}]{RN39}%
  \BibitemOpen
  \bibfield  {author} {\bibinfo {author} {\bibfnamefont {T.}~\bibnamefont
  {Clua-Provost}}, \bibinfo {author} {\bibfnamefont {A.}~\bibnamefont
  {Durand}}, \bibinfo {author} {\bibfnamefont {Z.}~\bibnamefont {Mu}}, \bibinfo
  {author} {\bibfnamefont {T.}~\bibnamefont {Rastoin}}, \bibinfo {author}
  {\bibfnamefont {J.}~\bibnamefont {Frauni\'e}}, \bibinfo {author}
  {\bibfnamefont {E.}~\bibnamefont {Janzen}}, \bibinfo {author} {\bibfnamefont
  {H.}~\bibnamefont {Schutte}}, \bibinfo {author} {\bibfnamefont {J.~H.}\
  \bibnamefont {Edgar}}, \bibinfo {author} {\bibfnamefont {G.}~\bibnamefont
  {Seine}}, \bibinfo {author} {\bibfnamefont {A.}~\bibnamefont {Claverie}},
  \bibinfo {author} {\bibfnamefont {X.}~\bibnamefont {Marie}}, \bibinfo
  {author} {\bibfnamefont {C.}~\bibnamefont {Robert}}, \bibinfo {author}
  {\bibfnamefont {B.}~\bibnamefont {Gil}}, \bibinfo {author} {\bibfnamefont
  {G.}~\bibnamefont {Cassabois}},\ and\ \bibinfo {author} {\bibfnamefont
  {V.}~\bibnamefont {Jacques}},\ }\bibfield  {title} {\bibinfo {title}
  {Isotopic control of the boron-vacancy spin defect in hexagonal boron
  nitride},\ }\href {https://doi.org/10.1103/PhysRevLett.131.126901} {\bibfield
   {journal} {\bibinfo  {journal} {Physical Review Letters}\ }\textbf {\bibinfo
  {volume} {131}},\ \bibinfo {pages} {126901} (\bibinfo {year}
  {2023})}\BibitemShut {NoStop}%
\bibitem [{\citenamefont {Cai}\ \emph {et~al.}(2013)\citenamefont {Cai},
  \citenamefont {Retzker}, \citenamefont {Jelezko},\ and\ \citenamefont
  {Plenio}}]{RN40}%
  \BibitemOpen
  \bibfield  {author} {\bibinfo {author} {\bibfnamefont {J.}~\bibnamefont
  {Cai}}, \bibinfo {author} {\bibfnamefont {A.}~\bibnamefont {Retzker}},
  \bibinfo {author} {\bibfnamefont {F.}~\bibnamefont {Jelezko}},\ and\ \bibinfo
  {author} {\bibfnamefont {M.~B.}\ \bibnamefont {Plenio}},\ }\bibfield  {title}
  {\bibinfo {title} {A large-scale quantum simulator on a diamond surface at
  room temperature},\ }\href
  {https://doi.org/https://doi.org/10.1038/nphys2519} {\bibfield  {journal}
  {\bibinfo  {journal} {Nature Physics}\ }\textbf {\bibinfo {volume} {9}},\
  \bibinfo {pages} {168} (\bibinfo {year} {2013})}\BibitemShut {NoStop}%
\bibitem [{\citenamefont {Aslam}\ \emph {et~al.}(2017)\citenamefont {Aslam},
  \citenamefont {Pfender}, \citenamefont {Neumann}, \citenamefont {Reuter},
  \citenamefont {Zappe}, \citenamefont {Fávaro~de Oliveira}, \citenamefont
  {Denisenko}, \citenamefont {Sumiya}, \citenamefont {Onoda}, \citenamefont
  {Isoya},\ and\ \citenamefont {Wrachtrup}}]{RN41}%
  \BibitemOpen
  \bibfield  {author} {\bibinfo {author} {\bibfnamefont {N.}~\bibnamefont
  {Aslam}}, \bibinfo {author} {\bibfnamefont {M.}~\bibnamefont {Pfender}},
  \bibinfo {author} {\bibfnamefont {P.}~\bibnamefont {Neumann}}, \bibinfo
  {author} {\bibfnamefont {R.}~\bibnamefont {Reuter}}, \bibinfo {author}
  {\bibfnamefont {A.}~\bibnamefont {Zappe}}, \bibinfo {author} {\bibfnamefont
  {F.}~\bibnamefont {Fávaro~de Oliveira}}, \bibinfo {author} {\bibfnamefont
  {A.}~\bibnamefont {Denisenko}}, \bibinfo {author} {\bibfnamefont
  {H.}~\bibnamefont {Sumiya}}, \bibinfo {author} {\bibfnamefont
  {S.}~\bibnamefont {Onoda}}, \bibinfo {author} {\bibfnamefont
  {J.}~\bibnamefont {Isoya}},\ and\ \bibinfo {author} {\bibfnamefont
  {J.}~\bibnamefont {Wrachtrup}},\ }\bibfield  {title} {\bibinfo {title}
  {Nanoscale nuclear magnetic resonance with chemical resolution},\ }\href
  {https://doi.org/10.1126/science.aam8697} {\bibfield  {journal} {\bibinfo
  {journal} {Science}\ }\textbf {\bibinfo {volume} {357}},\ \bibinfo {pages}
  {67} (\bibinfo {year} {2017})}\BibitemShut {NoStop}%
\bibitem [{\citenamefont {Morishita}\ \emph {et~al.}(2023)\citenamefont
  {Morishita}, \citenamefont {Morioka}, \citenamefont {Nishikawa},
  \citenamefont {Yao}, \citenamefont {Onoda}, \citenamefont {Abe},
  \citenamefont {Ohshima},\ and\ \citenamefont {Mizuochi}}]{RN42}%
  \BibitemOpen
  \bibfield  {author} {\bibinfo {author} {\bibfnamefont {H.}~\bibnamefont
  {Morishita}}, \bibinfo {author} {\bibfnamefont {N.}~\bibnamefont {Morioka}},
  \bibinfo {author} {\bibfnamefont {T.}~\bibnamefont {Nishikawa}}, \bibinfo
  {author} {\bibfnamefont {H.}~\bibnamefont {Yao}}, \bibinfo {author}
  {\bibfnamefont {S.}~\bibnamefont {Onoda}}, \bibinfo {author} {\bibfnamefont
  {H.}~\bibnamefont {Abe}}, \bibinfo {author} {\bibfnamefont {T.}~\bibnamefont
  {Ohshima}},\ and\ \bibinfo {author} {\bibfnamefont {N.}~\bibnamefont
  {Mizuochi}},\ }\bibfield  {title} {\bibinfo {title} {Spin-dependent dynamics
  of photocarrier generation in electrically detected nitrogen-vacancy-based
  quantum sensing},\ }\href {https://doi.org/10.1103/PhysRevApplied.19.034061}
  {\bibfield  {journal} {\bibinfo  {journal} {Physical Review Applied}\
  }\textbf {\bibinfo {volume} {19}},\ \bibinfo {pages} {034061} (\bibinfo
  {year} {2023})}\BibitemShut {NoStop}%
\bibitem [{\citenamefont {Gao}\ \emph {et~al.}(2025)\citenamefont {Gao},
  \citenamefont {Vaidya}, \citenamefont {Li}, \citenamefont {Ge}, \citenamefont
  {Dikshit}, \citenamefont {Zhang}, \citenamefont {Ju}, \citenamefont {Shen},
  \citenamefont {Jin}, \citenamefont {Ping},\ and\ \citenamefont {Li}}]{RN43}%
  \BibitemOpen
  \bibfield  {author} {\bibinfo {author} {\bibfnamefont {X.}~\bibnamefont
  {Gao}}, \bibinfo {author} {\bibfnamefont {S.}~\bibnamefont {Vaidya}},
  \bibinfo {author} {\bibfnamefont {K.}~\bibnamefont {Li}}, \bibinfo {author}
  {\bibfnamefont {Z.}~\bibnamefont {Ge}}, \bibinfo {author} {\bibfnamefont
  {S.}~\bibnamefont {Dikshit}}, \bibinfo {author} {\bibfnamefont
  {S.}~\bibnamefont {Zhang}}, \bibinfo {author} {\bibfnamefont
  {P.}~\bibnamefont {Ju}}, \bibinfo {author} {\bibfnamefont {K.}~\bibnamefont
  {Shen}}, \bibinfo {author} {\bibfnamefont {Y.}~\bibnamefont {Jin}}, \bibinfo
  {author} {\bibfnamefont {Y.}~\bibnamefont {Ping}},\ and\ \bibinfo {author}
  {\bibfnamefont {T.}~\bibnamefont {Li}},\ }\bibfield  {title} {\bibinfo
  {title} {Single nuclear spin detection and control in a van der waals
  material},\ }\href {https://doi.org/10.1038/s41586-025-09258-7} {\bibfield
  {journal} {\bibinfo  {journal} {Nature}\ }\textbf {\bibinfo {volume} {643}},\
  \bibinfo {pages} {943–949} (\bibinfo {year} {2025})}\BibitemShut {NoStop}%
\bibitem [{\citenamefont {Singh}\ \emph {et~al.}(2025)\citenamefont {Singh},
  \citenamefont {Robertson}, \citenamefont {Scholten}, \citenamefont {Healey},
  \citenamefont {Abe}, \citenamefont {Ohshima}, \citenamefont {Tan},
  \citenamefont {Kianinia}, \citenamefont {Aharonovich}, \citenamefont
  {Broadway}, \citenamefont {Reineck},\ and\ \citenamefont {Tetienne}}]{RN44}%
  \BibitemOpen
  \bibfield  {author} {\bibinfo {author} {\bibfnamefont {P.}~\bibnamefont
  {Singh}}, \bibinfo {author} {\bibfnamefont {I.~O.}\ \bibnamefont
  {Robertson}}, \bibinfo {author} {\bibfnamefont {S.~C.}\ \bibnamefont
  {Scholten}}, \bibinfo {author} {\bibfnamefont {A.~J.}\ \bibnamefont
  {Healey}}, \bibinfo {author} {\bibfnamefont {H.}~\bibnamefont {Abe}},
  \bibinfo {author} {\bibfnamefont {T.}~\bibnamefont {Ohshima}}, \bibinfo
  {author} {\bibfnamefont {H.~H.}\ \bibnamefont {Tan}}, \bibinfo {author}
  {\bibfnamefont {M.}~\bibnamefont {Kianinia}}, \bibinfo {author}
  {\bibfnamefont {I.}~\bibnamefont {Aharonovich}}, \bibinfo {author}
  {\bibfnamefont {D.~A.}\ \bibnamefont {Broadway}}, \bibinfo {author}
  {\bibfnamefont {P.}~\bibnamefont {Reineck}},\ and\ \bibinfo {author}
  {\bibfnamefont {J.-P.}\ \bibnamefont {Tetienne}},\ }\bibfield  {title}
  {\bibinfo {title} {Violet to near-infrared optical addressing of spin pairs
  in hexagonal boron nitride},\ }\href {https://doi.org/10.1002/adma.202414846}
  {\bibfield  {journal} {\bibinfo  {journal} {Advanced Materials}\ }\textbf
  {\bibinfo {volume} {37}},\ \bibinfo {pages} {2414846} (\bibinfo {year}
  {2025})}\BibitemShut {NoStop}%
\end{thebibliography}
\end{document}

\title{Supplementary Material for \vspace{0.08cm}\\
    \small ``Room-Temperature Electrical Readout of Spin Defects in van der Waals Materials''}

    \author{{Shihao Ru$^{1,2,3,4}$, Liheng An$^2$, Haidong Liang$^{5}$, Zhengzhi Jiang$^6$, Zhiwei Li$^{1}$, Xiaodan Lyu$^{2}$, Feifei Zhou$^{7}$, Hongbing Cai$^{8}$, Yuzhe Yang$^{2}$, Ruihua He$^{2,9}$, Robert Cernansky$^{10}$, Edwin Hang Tong Teo$^{1}$, Manas Mukherjee$^{11}$,  Andrew A. Bettiol$^{5}$, Jesus Z\'u\~niga-Perez$^{2, 12}$, Fedor Jelezko$^{10, *}$, and Weibo Gao$^{1, 2, 3, 4, \dagger}$\vspace{0.2cm}}
    \textit{\small \newline$^1$School of Electrical and Electronic Engineering, Nanyang Technological University, Singapore\\
    $^2$Division of Physics and Applied Physics, School of Physical and Mathematical Sciences, Nanyang Technological University, Singapore 637371, Singapore\\
    $^3$Centre for Quantum Technologies, Nanyang Technological University, Singapore 117543, Singapore\\
    $^4$National Centre for Advanced Integrated Photonics, Nanyang Technological University, Singapore 639798, Singapore\\
    $^5$Centre for Ion Beam Applications, Department of Physics, National University of Singapore, Singapore 117542, Singapore\\
    $^6$Department of Chemistry, National University of Singapore, Singapore 117543, Singapore\\
    $^{7}$College of Metrology Measurement and Instrument, China Jiliang University, Hangzhou, 310018, China\\
    $^8$Hefei National Laboratory for Physical Sciences at the Microscale, University of Science and Technology of China, Hefei Anhui 230026, China\\
    $^9$School of Biological Sciences, Nanyang Technological University, Singapore 637551, Singapore\\
    $^10$Institute for Quantum Optics, Ulm University, Albert-Einstein-Allee 11, 89081 Ulm, Germany\\
    $^{11}$Centre for Quantum Technologies, National University of Singapore, Singapore 117543, Singapore\\
    $^{12}$Majulab, International Research Laboratory IRL 3654, CNRS, Université Côte d’Azur, Sorbonne Université, National University of Singapore, Nanyang Technological University, Singapore, Singapore
    }}

\date{\today}

\maketitle
  
\renewcommand{\theequation}{S\arabic{equation}}
\renewcommand{\thefigure}{S\arabic{figure}}

In this Supplementary Information, we provide further information on:\vspace{0.1cm}\\
  \hspace*{0.6cm}Section 1:  Sample preparation\\
  \hspace*{0.6cm}Section 2:  Experimental Setup, magnetic field alignment and microwave antenna\\
  \hspace*{0.6cm}Section 3:  Optical images, AFM results and photocurrent results\\
  \hspace*{0.6cm}Section 4:  XY-8 measurements of thin hBN sample\\
    \hspace*{0.6cm}Section 5:  PDMR results of thick hBN sample\\
  \hspace*{0.6cm}Section 6:  Supplementary Information Figures\vspace{0.1cm}
  
  All figures and equations in Supplementary Information are labeled with the prefix ``S'' to distinguish from those in the manuscript.

\section{Section I: Sample preparation}
For the sample measured in the main text, electrodes consisting of 1 nm Cr and 40 nm Au were first fabricated on a SiO$_2$/Si substrate using a standard electron beam lithography (EBL) process.
We employed high-pressure, high-temperature–grown hBN bulk crystals obtained from the National Institute for Materials Science (Japan) in this work.
Flakes were exfoliated from these high-quality single crystals using the Scotch tape method. The tape was then pressed onto a 285 nm SiO$_2$ substrate and left in place for one day to improve flake yield.
A thin hBN flake was then transferred onto the central region of the electrode using a polydimethylsiloxane (PDMS) and polycarbonate (PC) stamping method.
The hBN used in our experiments had a thickness of approximately 18 nm (measured by AFM, about 50 layers).
Then we used a focused ion beam (FIB, Raith Velion) for VB defects generation ($\sim 200$ ppm), the hBN thin flake was subsequently implanted with Si$^+$ ions at an energy of 35 keV and a beam current of 6 pA, with an implantation dose of approximately 30 $\mu$C/cm$^2$.
Finally, a top thin few layers of graphite flake were transferred onto the hBN using the same standard transfer technique. After each transfer step, the sample was baked at 200 $^{\circ}$C for 5 minutes and then cleaned sequentially with chloroform, acetone, and isopropanol to remove the residual PC layer.

\section{Section II: Experimental Setup, magnetic field alignment and microwave antenna}
\begin{figure}[htpt]
  \centering
  \includegraphics[width=0.8\linewidth]{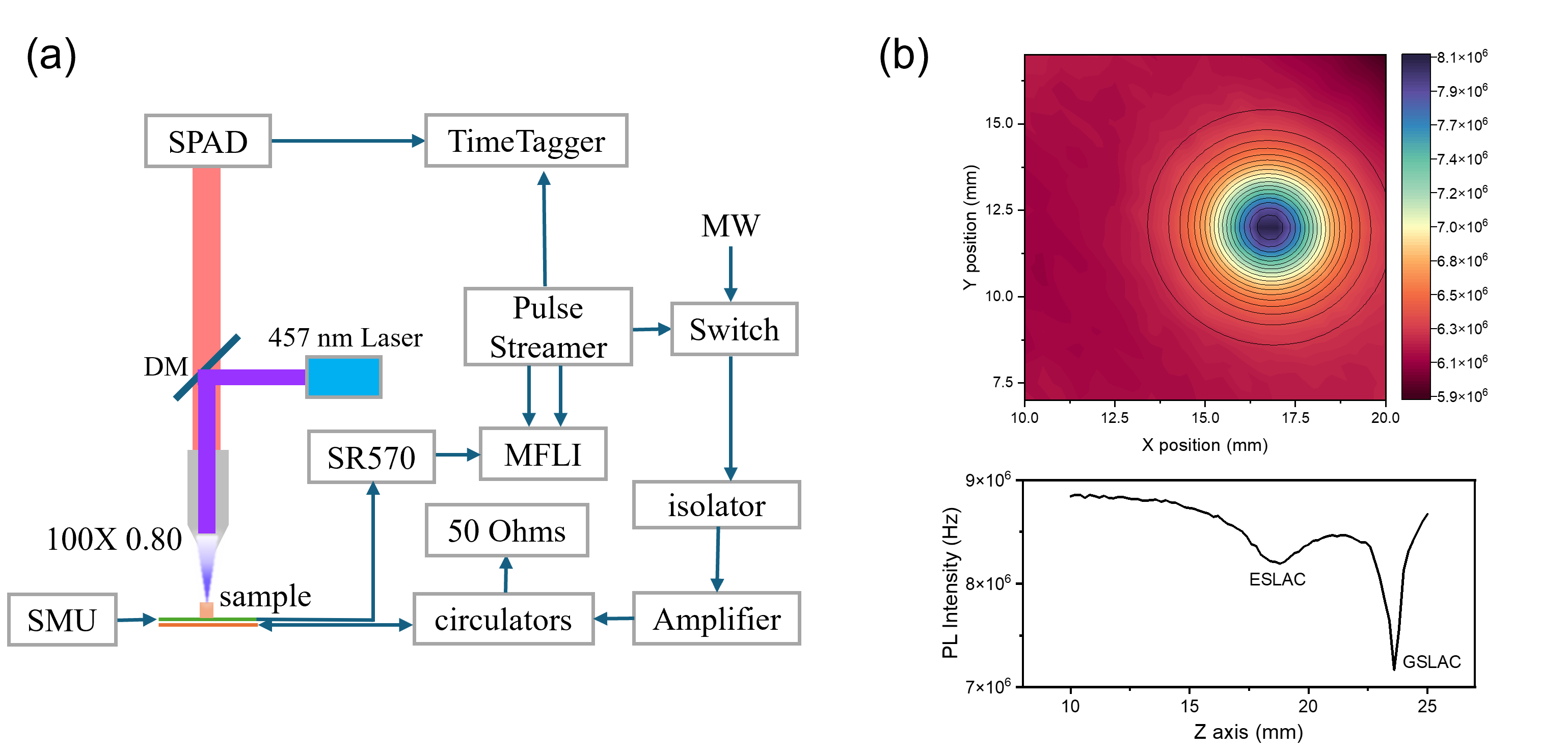}
  \caption{(a) Experimental setup. (b) Alignment of magnetic field orientation. In the top panel, the $z$-axis is fixed at the excited-state level anti-crossing (ESLAC) point of the $\mathrm{V_B^-}$ defects, and photoluminescence (PL) mapping is performed along the $x-$ and $y-$axes. This configuration is used throughout all experiments. We fitted the mapping results with a 2D Gauss function. In the bottom panel, the $x-$ and $y-$axes are fixed at the center positions obtained from fitting the top panel, and PL mapping is performed along the $z-$axis.}\label{Sfig1}
\end{figure}

In this section, we detail our measurement setup, with accompanying figures (Fig. \ref{Sfig1}). We also describe how a 3D stage and a permanent magnet are used to calibrate and align the external magnetic field with the BN crystal lattice. Finally, we introduce the microwave antenna used in the experiment.\\

All spin-control measurements were conducted using a home-built setup. The treated sample is mounted on a custom-designed PCB board and secured on an XYZ manual stage (Newport M-562F-XYZ) and an XYZ piezo scanner (PI P-611.3S), which is controlled by a digital piezo controller (PI E-727.3SD) for precise alignment of $\mathrm{V_B^-}$ defects with a microscope objective (Nikon CFI60 100X/0.80). Wire bonding was used to connect the electrodes and the PCB board.\\

A Cobolt 06-MLD-457-400 mW laser (HÜBNER Photonics) is used as the pump source, with direct modulation enabled by a time-to-live (TTL) signal generated by a digital signal generator (Pulse Streamer 8/2, Swabian Instruments). The photoluminescence (PL) emission is filtered from the excitation laser using a long-pass dichroic mirror (Thorlabs DMLP567R) and a long-pass filter (Thorlabs FELH0750), collected by Ø50 µm and 0.22 NA multi-mode fiber and subsequently detected by a single-photon avalanche diode (SPAD, SPCM-AQRH-44, Excelitas Technologies). The resulting photon arrival signals, along with the trigger output from the digital signal generator, are recorded by a streaming time-to-digital converter (Time Tagger Ultra, Swabian Instruments) for optically detected magnetic resonance measurements.\\

\begin{figure}[htpt]
  \centering
  \includegraphics[width=0.6\linewidth]{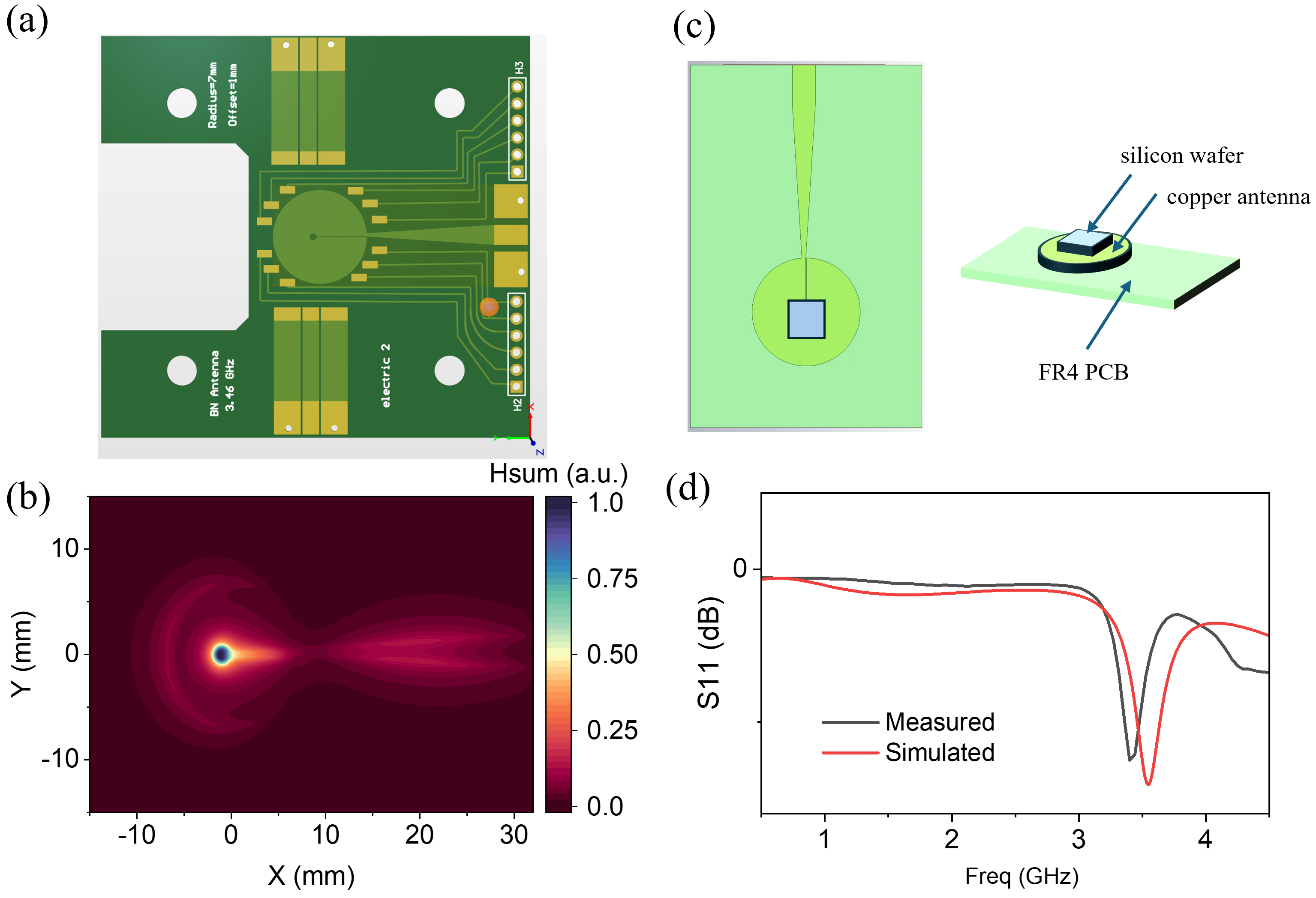}
  \caption{(a) Printed circuit board (PCB) board used in our experiment. The central area contains the antenna for applying microwaves to drive spin defects, the upper and lower electrodes are for connecting RF wire antenna, and the small electrodes near the central region are used for applying bias voltage and reading out the photocurrent signal. 
  (b) Simulated antenna structure designed using CST Studio software. From top to bottom, the sample consists of a silicon wafer, a copper antenna structure, and an FR4 PCB board. 
  (c) the simulated microwave magnetic field distribution on the silicon wafer top surface generated by the PCB antenna. 
  (d) Comparison between simulated and measured S11 parameters (microwave return loss) of this PCB antenna. The measurement was implemented by using a vector network analyzer (Rohde \& Schwarz). }\label{Sfig2}
\end{figure}

The microwave is generated by mixing a high-frequency signal from a Stanford Research Systems SG384 with a 500 MHz phase-modulated signal from the Arbitrary Waveform Generator (AWG70002B). To avoid noise from some sideband microwaves, the resulting microwave signal is passed through a high-pass filter and a low-pass filter, amplified and transmitted to a microwave antenna to serve as the driving field for $\mathrm{V_B^-}$ defects. The design and simulation results of the microwave antenna are shown in Fig. \ref{Sfig2}.\\

All TTL signals of spin-control sequences are precisely programmed and executed using the Pulse Streamer 8/2. Time-resolved optical signals are simultaneously recorded with the Time Tagger Ultra, ensuring accurate synchronization between spin manipulation and detection. For photoelectrically detected magnetic resonance measurements, the bias voltage is applied via the top graphite layer using a source meter unit (SMU, Keithley 2636B). The resulting photocurrent from the bottom electrode is amplified by a low-noise current preamplifier (SR570, Stanford Research Systems) and subsequently demodulated by a lock-in amplifier (MFLI, Zurich Instruments). The SR570 is configured with a sensitivity setting of 10 nA/V, corresponding to a transimpedance gain of 10$^8$ V/A. Both reference signals for lock-in demodulation are generated by the Pulse Streamer 8/2 to maintain synchronized and stable operation. The period of reference signal 1 is matched to the duration of the timing sequence used for measurement, thereby isolating the effective photoelectric signal. For signal acquisition, the lock-in time constant is set to 300 ms, and the output is averaged over a 1-second buffer to improve the signal-to-noise ratio. All measurements, including real-space mapping and time-domain pulse sequence experiments, are automated using a Python-based data acquisition framework.\\

To align the magnetic field with the symmetry axis of the $\mathrm{V_B^-}$ defects, a permanent magnet is mounted on a three-axis motorized translation stage (Thorlabs PT3/M-Z9), allowing fine control along the $x$, $y$ and $z$ directions. The alignment procedure leverages the sensitivity of the non-resonant PL signal to magnetic field orientation, as established in previous work \cite{Mu2022_prl,RN37}. This method enables precise calibration of the out-of-plane (z-axis) magnetic field by treating the process as a local optimization problem, as shown in Fig. \ref{Sfig1}(b).\\

The procedure involves three main steps:
\begin{itemize}
\item[(a)] The $y$-axis is first set to an estimated value, and a PL scan is performed along the $z$-axis for different $x$-axis positions to identify a local maximum.
\item[(b)] The $x$-axis is then fixed at the locally optimal position, and another $z$-axis PL scan is conducted while varying the y-axis position.
\item[(c)] Once the $z$-axis is adjusted to reach the excited-state level anti-crossing (ESLAC) point of the defects, a final 2D PL map is acquired across the $x-y$ plane to confirm alignment. We can use a 2D Gauss function $f\left(x,y\right)=z_0+A\exp\left[{-{\frac{1}{2}\left(\frac{x-x_c}{\omega_1}\right)}^2-{\frac{1}{2}\left(\frac{y-y_c}{\omega_2}\right)}^2}\right]$ to obtain the final $x_c$ and $y_c$ positions.
With a magnetic field orientation misalignment below 0.5$^{\circ}$, this approach ensures the magnetic field is well-aligned with the defect axis throughout the experiments.
\end{itemize}

Then we describe the custom-designed printed circuit board (PCB) used in our experiment, as shown in Fig. \ref{Sfig2}(a).
At the center of the board, we place the microwave antenna used to drive spin transitions in the defect ensemble. The top and bottom electrodes are used to apply radio frequency for nuclear spin driving.
The small electrodes near the central region are used to apply bias voltage and collect the resulting photocurrent. Microwave field distribution and related parameters were simulated using CST Studio. The antenna structure, illustrated in Fig. \ref{Sfig2}(b), is composed of three layers from top to bottom: a silicon wafer, a patterned copper antenna, and an FR4 PCB substrate (1.5 mm).
The hBN sample, mounted on a 500 $\mu$m-thick silicon wafer with a 285 nm SiO$_2$ layer, is placed at the central ring of the antenna. The simulated microwave magnetic field distribution at this position is shown in Fig. \ref{Sfig2}(c).
We also evaluated the microwave return loss (S11 parameter) of the antenna both by simulation and experimental measurement, as presented in Fig. \ref{Sfig2}(d). For radio frequency (RF) driving of nuclear transitions, a separate wire antenna was placed in close proximity to the sample.

\section{Section III: Optical images, AFM results and photocurrent results}

\begin{figure}[htpt]
  \centering
  \includegraphics[width=0.5\linewidth]{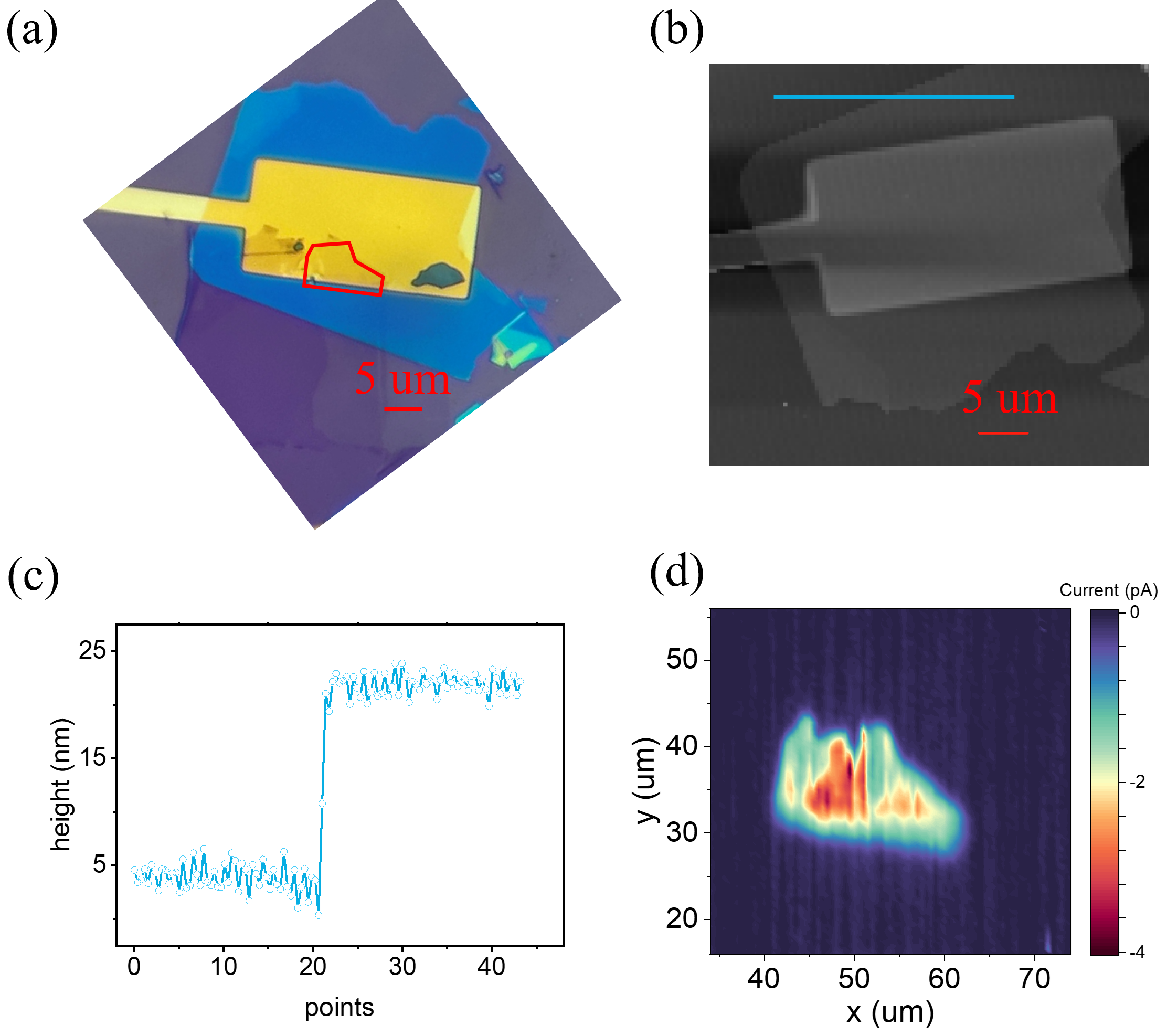}
  \caption{(a) Optical microscope image of the sample ith a red line indicating a 5-um reference length}. (b) AFM tomography with a red line indicating a 5-um reference length. (c) Height profile corresponding to the blue line in panel (b). (d) Photocurrent mapping results acquired under $-2.5$ V bias voltage and 2 mW laser illumination.\label{Sfig3}
\end{figure}

This part describes the sample device presented in the main text, the AFM results, and the variation in photocurrent. Fig. \ref{Sfig3}(a) shows the coloured image of the device under an optical microscope. From top to bottom, the layers consist of trilayer graphite, hBN, electrodes, and the silicon substrate.
A low-doping silicon wafer with a 285 nm SiO$_2$ layer on the surface was used. The device fabrication process is described in the Methods section of the main text. Fig. \ref{Sfig3}(b) shows the atomic force microscopy (AFM, Park NX10) tomography results.
Fig. \ref{Sfig3}(c) displays the profile across the blue line shown in Fig. \ref{Sfig3}(b). To quantify the height difference observed in the sample profile, we extracted two plateau regions from the height versus position data—one before and one after the step transition. The mean heights of these two regions were computed to estimate the height step, while the standard deviations were used to evaluate the uncertainty. The measured height of the thin hBN sample is $17.9 \pm 1.6$ nm.\\

\begin{figure}[htpt]
  \centering
  \includegraphics[width=0.4\linewidth]{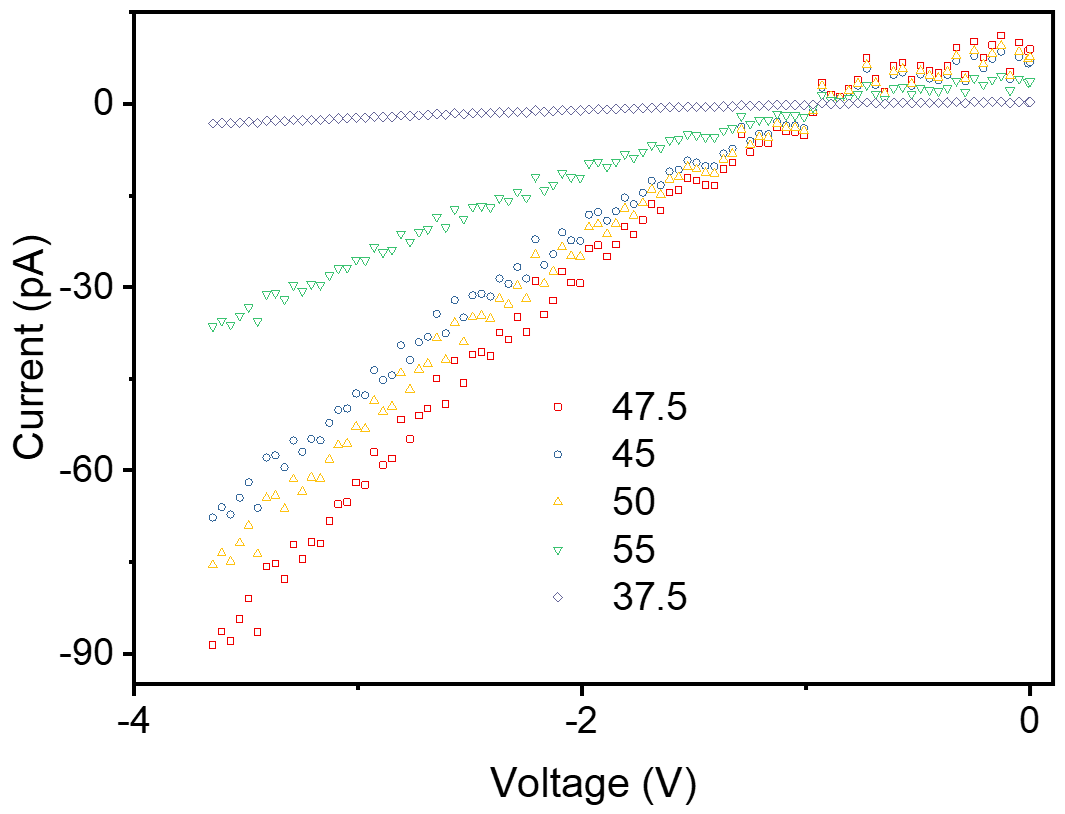}
  \caption{Measured photocurrent as a function of applied bias voltage under 2 mW laser excitation for different points. The legend values correspond to the $x$ location in Fig. \ref{Sfig3}(d) for a constant $y$ value of 35 $\mu$m.}\label{Sfig4}
\end{figure}

To determine whether this device is suitable for PDMR experiments, we first measured its leakage current behaviour. By applying 2 mW laser irradiation, we measured the device’s leakage current at $-2.5$ V.
The 2D mapping along the $x$ and $y$ axes is shown in Fig. \ref{Sfig3}(d).
A PI P-611.3S piezo scanner was used to perform three-dimensional scanning of the sample. The region with higher leakage current corresponds to the area outlined by the red dashed line in Fig. \ref{Sfig3}(a).
We measured the photocurrent variations under different bias voltages at various points along y = 35 $\mu$m in Fig \ref{Sfig4}, using a 2-mW laser. All measured results in the main text and results in S-Fig. 4 are correspond point $x = 47.5$ $\mu$m and $y = 35$ $\mu$m of Fig. \ref{Sfig3}(d).\\

To demonstrate the generality of our experiments, all measurements were carried out under ambient conditions (room temperature and air). Under these conditions, applying excessively high bias voltages can induce local breakdown and even burning of the graphite thin layer, which compromises device stability. Consequently, we selected a moderate bias voltage ($-2.3$ V) for the subsequent long-term spin measurements, which can provide good PDMR signals and signal-to-noise ratio while ensuring stable device operation. We also emphasize that all data presented in the main text were obtained from the same sample, which remained stable throughout the experiments.\\

\begin{figure}[htpt]
  \centering
  \includegraphics[width=0.8\linewidth]{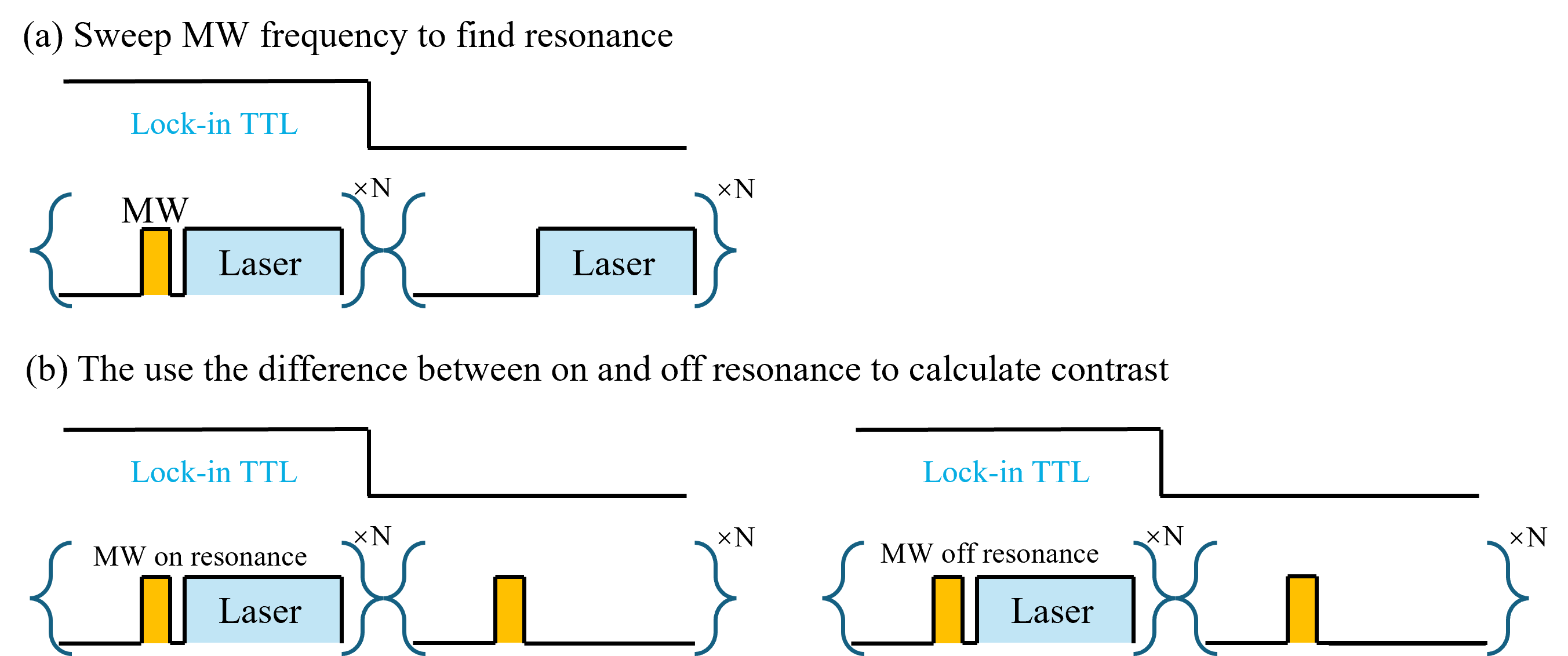}
  \caption{Timing sequence for pulsed PDMR. (a) The signal demodulated only captures the microwave induced photocurrent. (b) We measured first the photocurrent signal with microwave frequency on-resonance and then with microwave frequency off-resonance, which enabled to determine the magnetic resonance contrast.}\label{Sfig6}
\end{figure}

\begin{figure}[htpt]
  \centering
  \includegraphics[width=0.7\linewidth]{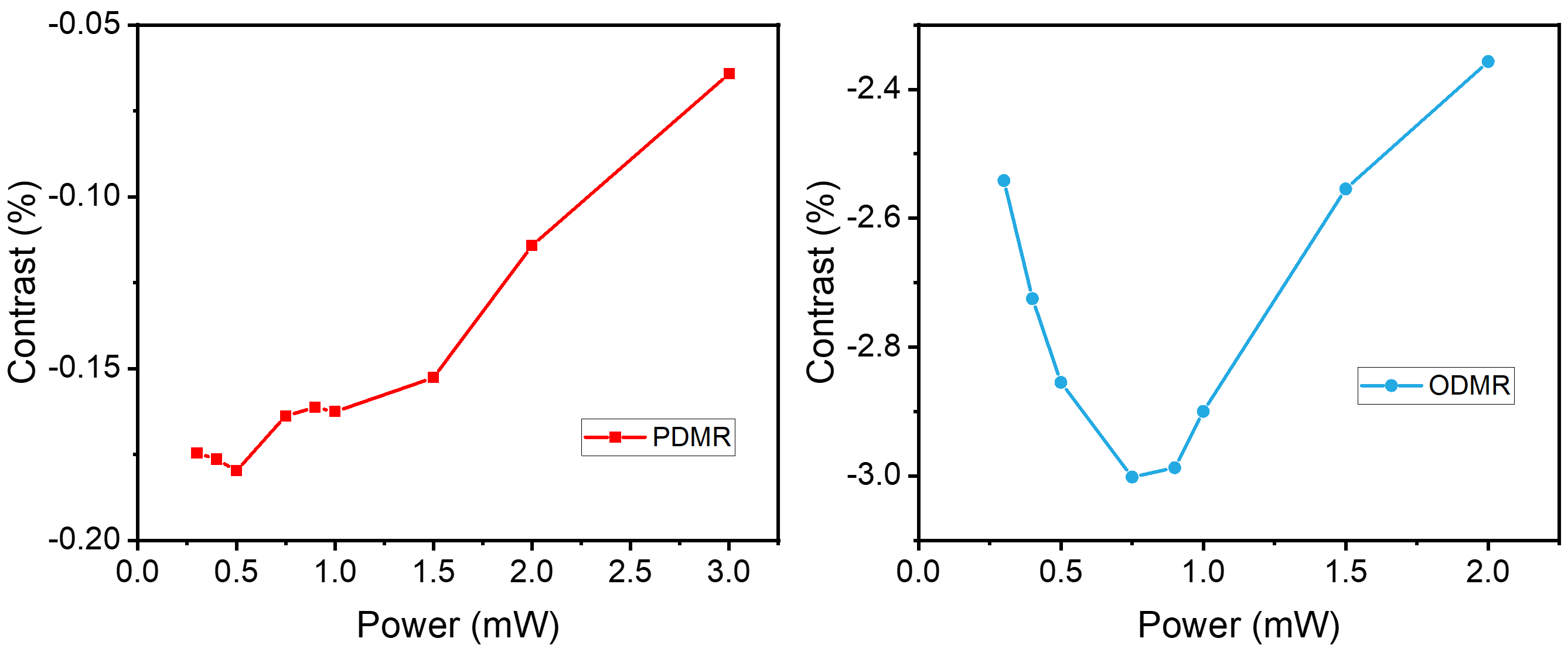}
  \caption{PDMR contrast (red squares) and ODMR contrast (blue circles) as functions of laser power at 2.3 V bias.}\label{Sfig5}
\end{figure}

The calculated method of PDMR contrasts is shown in Fig. \ref{Sfig6} \cite{RN26}. The resonance frequency is first identified by sweeping the microwave, as shown in Fig. \ref{Sfig6}(a). We also use this signal to calculate the SNR, as the sequence corresponds to the MW on/off condition.
Then for contrast measurements, the photocurrents are measured under on-resonance (signal) and off-resonance (reference) conditions (Fig. \ref{Sfig6}(b)).
Because both photocurrents share the same envelope frequency, the effect of the current-to-voltage preamplifier's bandwidth is identical in both cases.
Therefore, the contrast can be defined as (signal/reference$-$1). Fig. \ref{Sfig5} is a magnified view of Fig. 2(d) result in the manuscript.\\

For the SNR calculation, we define SNR as the effective photocurrent divided by the standard deviation of the background region and used in this work. It is important to highlight the difference in laser power saturation between ODMR and PDMR (Fig. 2(d)). With increasing laser power, the probability of photon re-excitation via two-photon absorption rises, thereby enhancing the effective photocurrent signal intensity and consequently increasing the SNR. The optical readout based on fluorescence follows the saturation of the fluorescence signals since the saturated emission rate is limited by the spontaneous decay rate of the excited state. In contrast, photoelectric readout is based on absorption leading to photo-ionisation and does not follow the same saturation behaviour. Therefore, at high excitation intensity, the number of photoelectrons and holes can exceed the number of emitted photons, resulting in a higher signal.

\section{Section IV: XY-8 measurements of thin hBN sample}

\begin{figure}[htpt]
  \centering
  \includegraphics[width=0.8\linewidth]{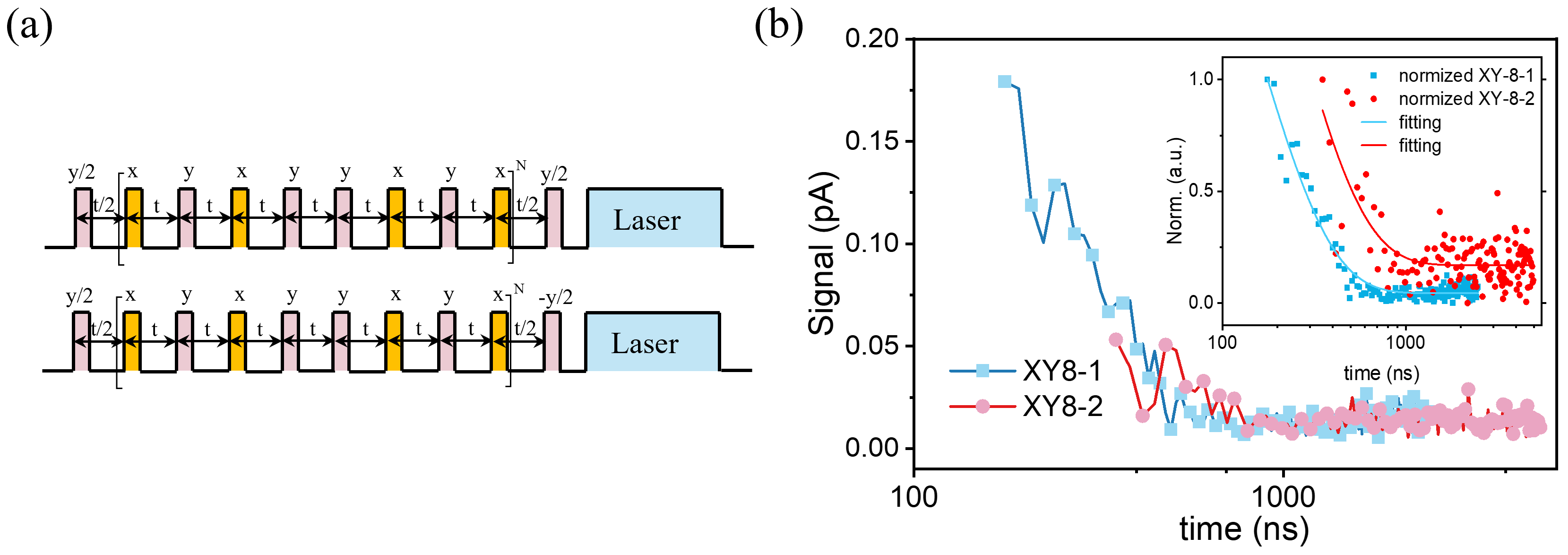}
  \caption{(a) Pulse sequence of the XY-8-N measurement. (b) Experimental results of the XY-8-N measurement. Due to spin locking and strong coupling with surrounding nitrogen nuclear spins, the performance of XY-8 is not as effective as that of CPMG. Inset figures are the normalized results with exp decay fitting.}\label{Sfig7}
\end{figure}

The results for CPMG dynamical decoupling have been presented in the main text.
Here, we report the measured results for the XY-8 dynamical decoupling sequence.
The XY-8 sequence, illustrated in Fig. \ref{Sfig7}(a), consists of two main components: the signal and reference sequences.
The result is obtained through lock-in detection by subtracting the reference sequence from the signal sequence, effectively eliminating common-mode noise and enhancing photocurrent signals.
Spin states are initialized and read out using laser pulses.
A $\pi/2$ pulse along the $y$-axis is applied at the beginning to prepare the spin in a superposition state, and another $\pi/2$ pulse at the end serves for readout.
In between, an N-set of eight $\pi$ pulses is applied with alternating phases following the pattern: $x–y–x–y–y–x–y–x$, which serves to extend the coherence time.
This alternating phase arrangement effectively suppresses pulse errors to some extent. In the reference sequence, the final $\pi/2$ pulse is replaced by a $-\pi/2$ pulse along the $y$-axis. The measured XY-8 results are shown in Fig. \ref{Sfig7}(b).
As discussed in the main text, the XY-8 protocol is less effective at preserving the coherence time for $\mathrm{V_B^-}$ centers in hBN.
For $N = 1$, the coherence time measured in XY-8 sequence is $144.56 \pm 6.10$ ns as we used $y =y_0+A_0\exp(-t/t_0)$ to fit it. This reduced performance may arise from spin-locking effects and strong coupling to nearby nitrogen nuclear spins, 
whereas CPMG sequences are less sensitive to pulse errors and thus maintain coherence more effectively.

\section{Section VI: PDMR results of thick hBN sample}

\begin{figure}[htpt]
  \centering
  \includegraphics[width=0.75\linewidth]{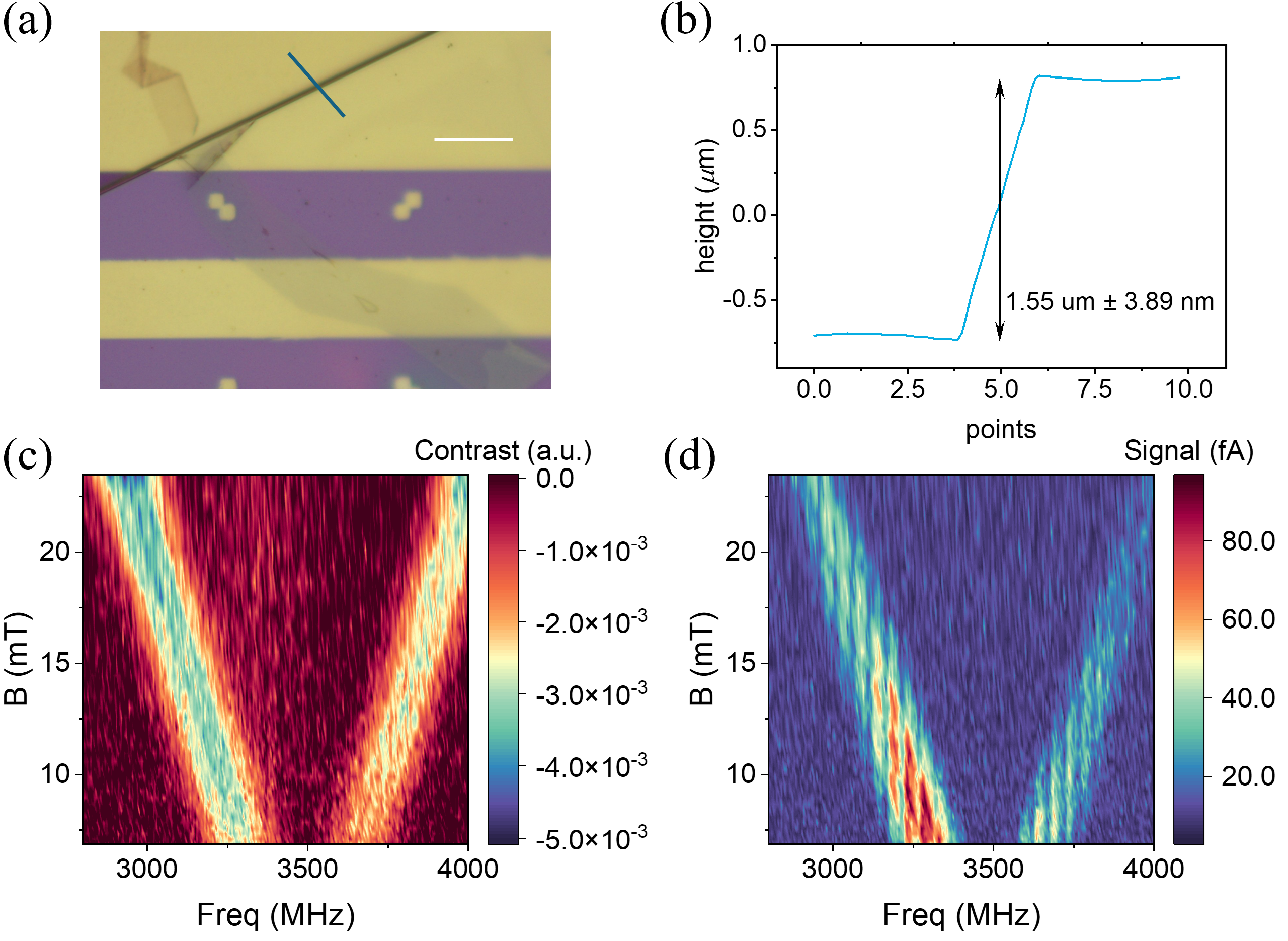}
  \caption{Results for sample II. (a) Optical microscope image of the thick sample with a white line indicating a 20-um reference length. (b) Height AFM profile corresponding to the blue line in panel (a). (c) and (d) are magnetic field dependent ODMR and PDMR spectra of $m_s=\pm1$ at bias voltage 20 V.}\label{Sfig8}
\end{figure}

We also fabricated a similar structure using thick hBN. Fig. \ref{Sfig8}(a) shows an optical image of the thick hBN device used in the PDMR experiments, with a white line indicating a 20 $\nu$m reference length.
The strip-shaped top graphite layer applies a bias voltage to the sample through the upper electrode shown in Fig. \ref{Sfig8}(a), while the 20 $\nu$m-wide electrodes at the bottom of Fig. \ref{Sfig8}(a) is used to measure the photocurrent signal.
An AFM scan was performed on the hBN used in this device, revealing a thickness of $1550 \pm 4$ nm, as shown in Fig. \ref{Sfig8}(b).
The increased thickness of the hBN allows for the application of higher bias voltages. We also performed measurements on this sample. Fig. \ref{Sfig8}(c) and \ref{Sfig8}(d) show the pulsed ODMR and PDMR results under varying magnetic fields, respectively.
For this sample, the pulsed PDMR signal reaches a maximum of only 0.1 pA, which is significantly lower than that of the thin hBN sample discussed in the main text.
The reduced signal may be due to the presence of more unintended defects in thick hBN, which can trap carriers and hinder their transport to the measurement electrodes.
The pulsed PDMR results for the thick hBN device were obtained using the same measurement setup as for the thin hBN device; but due to the higher optical signal, an amplified photodetector was used instead of a SPAD for the pulsed PDMR measurements.

\clearpage
\newpage
\section{Section VII: Supplementary Information Figures}

\clearpage
\newpage
\nocite{*}
% \bibliography{sup_bib.bib}